%% file: BHTpartV.tex
\documentclass[11pt,a4paper]{article}
\pdfoutput=1
\usepackage{jheppub}

\usepackage{multirow}
\usepackage{natbib}
\usepackage{amsmath,amssymb,amsfonts}
\usepackage{graphicx}
\usepackage{color}
\usepackage{bbm}
\usepackage{url}
\usepackage{slashed}
\usepackage{subfigure}
\usepackage[usenames,dvipsnames]{xcolor}
\usepackage{float} 
\usepackage{epsfig}
\usepackage{adjustbox}
\usepackage{graphicx}
\usepackage{placeins}
\usepackage{euscript}
\usepackage{color}
\usepackage{slashed}
\usepackage{epstopdf}
\usepackage{mathbbol}
\usepackage{mathrsfs}
\usepackage[normalem]{ulem}
\usepackage{pifont}
\usepackage{bbold}
\usepackage{float}
\allowdisplaybreaks
\usepackage{calc}

\usepackage{accents}

\usepackage{bbding}

\usepackage{verbatim}
\usepackage{graphicx}
\usepackage{latexsym}

\newcommand{\Eqref}[1]{Eq.~\eqref{#1}}

\newcommand{\be}{\begin{equation}}
\newcommand{\ee}{\end{equation}}
\def \eea{\end{eqnarray}}
\def \bea{\begin{eqnarray}}

\definecolor{darkgreen}{rgb}{0.2,0.6,0}
\definecolor{lightblue}{rgb}{0,0.5,0.8}
\definecolor{lightred}{rgb}{0.8,0.2,0.2}
\definecolor{darkorange}{rgb}{1,0.549,0}
\definecolor{brown}{rgb}{0.609, 0.164, 0.164}

\allowdisplaybreaks

\title{Rotating black hole thermodynamics at NNLO in higher-curvature gravity}

\author[a]{Cristóbal Laporte} \emailAdd{calaporte@icc.ub.edu}
\affiliation[a]{Departament de Física Quantica i Astrofísica, Institut de Ci\`encies del Cosmos, Universitat de Barcelona, Martí i Franqu\`es 1, E-08028 Barcelona, Spain}

\abstract{
We investigate higher-curvature corrections to black-hole thermodynamics in quasitopological gravity using perturbative Euclidean-action methods. Static, spherically symmetric black holes provide a benchmark for the action expansion, while our main application concerns five-dimensional, asymptotically flat, rotating black holes with equal angular momenta. For Einstein gravity supplemented by Gauss--Bonnet and cubic quasitopological interactions, we compute the Gibbs free energy through next-to-next-to-leading order, retaining both quadratic and cubic corrections. The result retains the full dependence on the background spin parameter, without a slow-rotation expansion, and extends to arbitrary couplings of quadratic and cubic curvature invariants. We derive the corrected mass, entropy and angular momenta, and show how rotation affects the size and sign of the corrections. 
}

\begin{document}
\maketitle
\section{Introduction}
\label{sect.Intro}

From the point of view of effective field theory, the Einstein-Hilbert action is only the first term of an expansion in derivatives. Higher-curvature terms are generated by quantum corrections \cite{Donoghue:1994dn,Burgess:2003jk} and appear explicitly in string theory \cite{Zwiebach:1985uq,Gross:1986iv}, where their coefficients encode information about the ultraviolet completion of gravity. Black hole thermodynamics is a natural place to study its effects. The mass, entropy and angular momenta of a black hole, as functions of its temperature and angular velocities, are sensitive to the higher-derivative terms both through the corrected geometry and the Wald entropy \cite{Wald:1993nt,Iyer:1994ys,Jacobson:1993vj}. Corrections of this type have played an important role, for instance, in the relation between extremality bounds and the weak gravity conjecture \cite{Kats:2006xp,Cheung:2018cwt,Goon:2019faz} and in precision tests of holography \cite{Bobev:2020egg,Bobev:2022bjm,Cassani:2022lrk}.

The direct way of computing these corrections requires the corrected solution, and this is a serious obstacle for rotating black holes. Even in Einstein--Gaus--Bonnet gravity \cite{Lovelock:1971yv}, the simplest higher-curvature theory in five dimensions, no rotating black hole is known in closed form for generic values of the coupling. The available results rely on special values of the couplings \cite{Anabalon:2009kq}, slow rotation \cite{Kim:2007iw,Fierro:2020wps}, numerical methods \cite{Brihaye:2008kh,Brihaye:2010wx,Konoplya:2020fbx} or expansions in the coupling \cite{Ma:2020xwi,Ma:2025vjk}. The situation is similar in four dimensions, where the corrections to the Kerr metric have been obtained as expansions in spin \cite{Cardoso:2018ptl,Cano:2019ore} or numerically \cite{Fernandes:2025vxg}. 

A substantial simplification follows from an observation by Reall and Santos \cite{Reall:2019sah}. The leading correction to the thermodynamic potential can be obtained by evaluating the higher-derivative terms of the action on the uncorrected solution, provided that the comparison is made at fixed temperature and angular velocities. This allowed them to compute the corrections to the thermodynamics of the Kerr black hole analytically, and the method has since been applied and refined in a variety of settings; see \cite{Ma:2023qqj,Hu:2023gru,Ma:2024ynp,Wu:2024iiz,Ma:2024ulp,Hu:2025ogz}. For rotating black holes in higher dimensions, the leading quadratic curvature corrections to the thermodynamics of Myers-Perry black holes \cite{Myers:1986un} with general rotation parameters were obtained in \cite{Wu:2024iiz}. Rotating AdS and Taub--NUT black holes were studied in \cite{Guo:2025muo,Chen:2025ary,Lu:2026clu,Chen:2024knw,Chen:2026bky}, in some cases including cubic terms at leading order. 

The prescription can be extended beyond leading order. It was shown in \cite{Ma:2023qqj} that the Euclidean action of asymptotically flat black holes at next-to-next-to-leading order (NNLO) requires the solution only at next-to-leading order (NLO). There is a simple reason why this order is relevant in five dimensions. In four dimensions, the four derivative terms do not modify the vacuum solutions \cite{Lu:2015psa,Pravda:2016fue}, so the leading corrections come from six derivative terms \cite{Daas:2023axu,Daas:2024pxs,Daas:2025lzr} and are capture by the Reall-Santos prescription \cite{Reall:2019sah}. In five dimensions, the Gauss--Bonnet term is no longer topological and provides the leading correction. The cubic terms then enter at the same order as the square of the Gauss--Bonnet coupling, and a consistent treatment of the six-derivative order requires the first-order correction of the solution. To the best of our knowledge, no results beyond leading order are available for rotating black holes. This is the problem we address in this paper.

We consider five-dimensional Einstein gravity supplemented by the Gauss--Bonnet term and a cubic quasitopological interaction \cite{Oliva:2010eb,Myers:2010ru} and we compute the thermodynamics of asymptotically flat rotating black holes with equal angular momenta through NNLO, that is, including the terms linear and quadratic in the Gauss--Bonnet coupling and linear in the cubic one. Quasitopological gravities \cite{Oliva:2010eb,Myers:2010ru,Dehghani:2011vu,Ahmed:2017jod,Cisterna:2017umf,Bueno:2019ycr,Bueno:2022res,Moreno:2023rfl,Bueno:2025qjk,Murcia:2026pmg,Bueno:2026wse} are a convenient framework for this problem. Their static, spherically symmetric black holes are determined by an algebraic equation, so exact thermodynamic expressions are available and provide a benchmark for the perturbative expansion. Moreover, it has been argued that these theories provide a basis for the gravitational effective action in five and higher-dimensions \cite{Bueno:2019ltp,Bueno:2024dgm}, and infinite towers of quasitopological interactions can resolve the Schwarzschild singularity \cite{Bueno:2024dgm,Bueno:2024zsx,Frolov:2024hhe,Bueno:2024eig,Hennigar:2025yqm,Bueno:2025tli,Borissova:2026wmn,Borissova:2026rbi,Bueno:2026oyg,Borissova:2026prd,Colleaux:2026hat}, which makes it natural to ask how their black holes are affected by rotation. In any case, at the order considered here the choice of cubic density is not restrictive: on a Ricci flat five dimensional background, arbitrary algebraic interactions contribute through a single combination of couplings, as we discuss in Section \ref{sect.Concl}.

Our starting point is the Reall--Santos prescription and its extension beyond leading order, which we review in Section \ref{Sect:OverviewFRG} and write it in a compact form valid at any order in the derivative expansion. Two aspects of the method become essential in the rotating case. First, the expansion is performed at fixed temperature and angular velocities, whereas a perturbation constructed at fixed mass and angular momenta changes both, so the parameters of the background must be adjusted accordingly. Second, the displacement of the horizon changes the domain of integration of the higher-curvature terms. We test the procedure on static black holes, where the NNLO free energy, mass and entropy reproduce the expansion of the exact results of \cite{Myers:2010ru}. 

We then turn to the Myers-Perry black hole with equal angular momenta. The first order correction to the metric induced by the Gauss--Bonnet term was constructed in \cite{Ma:2020xwi}. We rederive it in a closed form, preserving its full dependence on the spin, and use it to evaluate the Euclidean action. The result is a closed expression for the Gibbs free energy through NNLO, valid for arbitrary values of the spin below extremality, from which the mass, entropy and angular momenta follow. We also obtain the entropy at fixed mass and angular momenta, which at leading order is fixed by the shift of the free energy \cite{Reall:2019sah,Goon:2019faz} but beyond that order receives an additional contribution from the redefinition of the parameters. As consistency checks, the Wald entropy computed on the corrected horizon agrees at first order with the entropy derived from the free energy, and the static limit reproduces the expansion of the exact Einstein--Gauss--Bonnet black hole \cite{Boulware:1985wk}.

These results show how rotation affects the size and sign of the corrections. When the solutions are compared at the same temperature and angular velocity, the leading Gauss--Bonnet correction to the mass changes sign at a finite value of the spin, and so does the cubic contribution. The entropy correction linear in the Gauss--Bonnet coupling vanishes in the static limit at fixed temperature, but it is nonzero for rotating black holes. The NNLO terms contain logarithms of the spin parameter, which have a classical origin and should not be confused with the logarithmic quantum corrections to the entropy \cite{Sen:2012dw}. Finally, the expansion is not uniform near extremality, where the parameter shifts needed to keep the temperature and angular velocity fixed become singular. The extremal limit of perturbative higher-derivative corrections is known to be subtle \cite{Ma:2020xwi,Horowitz:2023xyl,Mao:2023qxq,Cano:2023dyg,Wu:2024iiz}, and it requires a separate analysis of the near-horizon geometry.

The paper is organized as follows. In Section \ref{Sect:OverviewFRG} we review the Reall-Santos prescription and its extension beyond leading order. In Section \ref{sect.3} we apply it to static, spherically symmetric black holes and compare with the exact results. Section \ref{sect.4} contains the rotating case, which contains the first order correction to the metric, the conditions that keep the temperature and angular velocities fixed, the evaluation of the Euclidean action and the resulting thermodynamic quantities. We conclude in Section \ref{sect.Concl}. Appendix \ref{apen.C} contains the general first-order solution and the tests of the ansatz, Appendix \ref{apen.D} the radial integration of the Gauss--Bonnet variation, and Appendix \ref{app:wald-first-order} the computation of the Wald entropy at first order.

\section{Review of Reall-Santos method}
\label{Sect:OverviewFRG}

The Euclidean action provides a way to calculate black hole thermodynamics with less information than is needed to construct the corrected solution. We review the prescription of \cite{Reall:2019sah} and its extension to NNLO \cite{Ma:2023qqj}, paying particular attention to the thermodynamic quantities that must remain fixed during the expansion. These conditions will guide both the static calculations in Section 3 and the rotating application that follows.

\subsection{Fixed thermodynamic sources and the leading correction}

The Reall-Santos prescription determines the leading correction to a black-hole thermodynamics potential by evaluating the higher-derivative action on the unperturbed solution \cite{Reall:2019sah}. Its extension beyond leading order retains this simplification in a weaker form: the action can be obtained without constructing the corrected solution to the same perturbative order \cite{Ma:2023qqj}. In this section, we review the expansion, emphasizing the boundary conditions needed in the rotating application.

Introducing a bookkeeping parameter $\epsilon$, we can write 
\begin{equation}\label{IeandFields}
    I_E\left[\Psi;\epsilon\right] = \sum^N_{n=0} \epsilon^n \, I_{n}\left[\psi\right] \, , \,\,\,\,\,\, \Psi(\epsilon) = \Psi_{(0)} + \sum_{n\ge1} \epsilon^n \, \Psi_{(n)} \, .
\end{equation}
Here $\Psi=\{g_{\mu\nu},A_\mu,\phi_\mu,...\}$ denotes the field content, specialized below the metric. The couplings are included in $I_n$, while $\epsilon$ records their order and is set to unity upon expansion. In the derivative counting used here, $I_n$ contains operators with $2 (n+1)$ derivatives. Curvature squared, curvature cubed, curvature quartic and curvature quintic interactions therefore enter at order $\epsilon$, $\epsilon^2$, $\epsilon^3$ and $\epsilon^4$, respectively. We restrict our attention to the perturbative branch continuously connected to the two-derivative solution. 

The functionals in Eq.\eqref{IeandFields} include the boundary terms and regularization appropriate to a chosen variational problem. In particular, the Einstein contribution includes the Gibbons-Hawking-York term and the reference subtraction in the asymptotically flat setting \cite{Gibbons:1976ue}. The expansion is performed at fixed thermodynamic sources and with a regular Euclidean horizon. For a neutral static solution, the source is the inverse temperature $\beta$ \cite{Hawking:1975vcx}. For a rotating solution, the angular potentials must also remain fixed, and the corresponding thermodynamic potential is
\begin{equation}\label{FreeEnergyI}
    I^{on}_E = \beta \, G \, , \,\,\,\,\,\, G = M - TS - \sum_i \Omega_i \, J_i \, , 
\end{equation}
where the superscript \textit{on} is a shorthand notation for \textit{on-shell}. For the static uncharged solutions considered here, this reduces to the Helmholtz free energy. Fixing $M$ and $J_i$ instead defines a different comparison between solutions and will be implemented separately after deriving the thermodynamic observables. 

Since $I_0$ is stationary at $\Psi_{(0)}$, its first variation vanishes for perturbations preserving the boundary conditions, since it is proportional to the leading-order field equations. Consequently
\begin{equation}\label{EuclideanActionRS}
    I^{on}_E = I_0\left[\Psi_{(0)}\right] + \epsilon I_1 \left[\Psi_{(0)}\right] + \mathcal{O}\left(\epsilon^2\right) \, .
\end{equation}
The first-order change of the metric does not contribute to this order. Eq.\eqref{EuclideanActionRS} is the leading order result found in \cite{Reall:2019sah}.

\subsection{Generalization to beyond leading order contributions}

In the following, $\delta^p  I_{n}\left[\Psi_{(r_1)},...,\Psi_{(r_p)}\right]$ denotes the \textit{p}th functional variation of $I_n$ evaluated at $\Psi_{(0)}$. The positive integers ($r_1,\ldots, r_p$) label the perturbative orders of the field corrections appearing as arguments, each of which specifies a direction of variation. All variations are performed with the boundary sources held fixed. Expanding the action through second order gives 
\begin{equation}\label{IENNLO}
    I^{on}_E = I_0\left[\Psi_{(0)}\right] + \epsilon \, I_1\left[\Psi_{(0)}\right] + \epsilon^2 \left(I_2\left[\Psi_{(0)}\right] + \frac{1}{2} \delta^2 I_0\left[\Psi_{(1)},\Psi_{(1)}\right]\right) + \mathcal{O}(\epsilon^3) \, .
\end{equation}
The term $\delta I_0[\Psi_{(2)}]$ has vanished by background stationarity. The first-order equation, contracted with an arbitrary admissible variation $\eta$, is 
\begin{equation}\label{VariationLO}
    \delta^2 I_0 \left[\Psi_{(1)},\eta\right] + \delta I_1\left[\eta\right] = 0 \, .
\end{equation}
Taking $\eta=\Psi_{(1)}$ in \Eqref{VariationLO} yields
\begin{equation}\label{IENNLOgeneric}
    I_E^{\mathrm{on}}=I_0[\Psi_{(0)}]+\epsilon I_1[\Psi_{(0)}]+\epsilon^2\left(I_2[\Psi_{(0)}]+\frac12\delta I_1[\Psi_{(1)}]\right)+O(\epsilon^3).
\end{equation}
Thus, the NNLO action requires the first-order perturbation but no second-order metric corrections. The $1/2$ factor incorporates the quadratic response of the Einstein action; it is not an extra factor in the higher-curvature density. 

A compact extension follows by differentiating the on-shell action with respect to $\epsilon$ along the family of solutions $\Psi(\epsilon)$ while holding all sources fixed
\begin{equation}\label{OnShellDer}
    \frac{d I_E^{\mathrm{on}}}{d\epsilon}
    =\sum_{j=1}^{N}j \, \epsilon^{j-1} \, I_j[\Psi(\epsilon)] \, .
\end{equation}
The term proportional to $d\Psi/d\epsilon$ vanishes by the stationarity of the complete action. If $\left[\epsilon^n\right]$ denotes the extraction of the corresponding Taylor coefficient, \Eqref{OnShellDer} implies
\begin{equation}\label{ExpansionOnShell}
    \left[\epsilon^n\right] \, I^{on}_{E} = \frac{1}{n} \sum^n_{j=1} j \left[\epsilon^{n-j}\right] I_{j}\left[\Psi(\epsilon)\right] \, , \quad 1\leq n \leq N \, .
\end{equation}
\Eqref{ExpansionOnShell} provides a practical implementation in terms of Taylor-expanded insertions without an explicit functional differentiation. In particular, the next two coefficients are 
\begin{align}
    [\epsilon^3]I_E^{\mathrm{on}}={}&I_3[\Psi_{(0)}]
    +\frac23\delta I_2[\Psi_{(1)}]
    +\frac13\delta I_1[\Psi_{(2)}]
    +\frac16\delta^2 I_1[\Psi_{(1)},\Psi_{(1)}],
    \label{cubic-order}\\
    [\epsilon^4]I_E^{\mathrm{on}}={}&I_4[\Psi_{(0)}]
    +\frac34\delta I_3[\Psi_{(1)}]
    +\frac12\delta I_2[\Psi_{(2)}]
    +\frac14\delta^2 I_2[\Psi_{(1)},\Psi_{(1)}]
    \nonumber\\
    &+\frac14\delta I_1[\Psi_{(3)}]
    +\frac14\delta^2 I_1[\Psi_{(1)},\Psi_{(2)}]
    +\frac1{24}\delta^3 I_1[\Psi_{(1)},\Psi_{(1)},\Psi_{(1)}].
    \label{quartic-order}
\end{align}
These expressions give the coefficients through ten derivatives in the stated counting, including products of lower-order couplings. In this representation, the coefficient of order $\epsilon^n$ can be obtained from the field expansion through $\Psi_{(n-1)}$. For example, direct evaluation of an insertion gives
\begin{align}
    I_1[\Psi_{(0)}+\epsilon\Psi_{(1)}+\epsilon^2\Psi_{(2)}]
    ={}& I_1[\Psi_{(0)}]+\epsilon\delta I_1[\Psi_{(1)}]
    \nonumber\\
    &+\epsilon^2\left(
    \delta I_1[\Psi_{(2)}]
    +\frac12\delta^2I_1[\Psi_{(1)},\Psi_{(1)}]
    \right)+O(\epsilon^3).
 \label{Insertion-example}
\end{align}
Substituting these Taylor coefficients into \Eqref{ExpansionOnShell} reproduces the weights above. The corrected metric changes both the Einstein action and the higher-derivative terms. The factors $(j/n)$ incorporate the Einstein contribution, allowing the corrections to the full on-shell action to be computed from the higher-curvature insertions alone. 

\Eqref{IENNLOgeneric} applies in the grand canonical ensemble, with $T$ and $\Omega_i$ held fixed. A metric perturbation constructed at fixed mass and angular momenta generally changes these thermodynamic variables. The background parameters must therefore be adjusted to cancel these changes. This adjustment can be included either in the metric perturbation or directly in the variation of the integrated higher-curvature term. In addition, if the horizon radius changes in the coordinates used for integration, the resulting change in the lower limit of the radial integral must be included. This contribution follows from varying the integral itself and does not require an additional boundary action at the horizon. We include both effects explicitly for the rotating case.

\section{Static and spherically symmetric black holes}
\label{sect.3}

We first apply the method to static black holes, for which the simplicity of the equations of motion in quasitopological gravity allows a direct comparison with exact thermodynamic expressions. Cubic interactions with simple equations in spherical symmetry were constructed in \cite{Oliva:2010eb,Myers:2010ru}. Their black hole solutions are determined by an algebraic equation rather than a higher-order radial differential equation. Here we use the five-dimensional cubic theory as a benchmark for the NNLO calculation and as the static limit of the rotating system studied in Section \ref{sect.4}.

We consider the asymptotically flat branch of the theory with Lorentzian bulk action
\begin{equation}\label{RotatingAction}
 I_{\mathrm{bulk}}=\frac{1}{16\pi G_N}\int d^5x\,\sqrt{-g}
 \left(R+\alpha\chi_4+\lambda\mathcal Z_3\right) \, ,
\end{equation}
with $\chi_4=R_{\mu\nu\rho\sigma}R^{\mu\nu\rho\sigma}-4R_{\mu\nu}R^{\mu\nu}+R^2$ is the Gauss--Bonnet density and the cubic invariant is normalized as
\begin{equation}\label{Z3Rot}
\begin{split}
    \mathcal Z_3={}R_a{}^e{}_c{}^f R^{abcd}&R_{bedf}
    +\frac17\bigg(\frac{21}{8}R R_{abcd}R^{abcd}
    -9R^{ab}R_a{}^{cde}R_{bcde}
    +15R^{ab}R^{cd}R_{acbd}\\
    &\hspace{18mm}+18R_a{}^bR_b{}^cR_c{}^a
    -\frac{33}{2}R R_{ab}R^{ab}+\frac{15}{8}R^3\bigg).
\end{split}
\end{equation}
Here $[\alpha]=L^2$ and $[\lambda]=L^4$, with $\alpha=O(\epsilon)$ and $\lambda=O(\epsilon^2)$ in the perturbative counting. For the static, spherically symmetric ansatz 
\begin{equation}
    ds^2 = -N(r)^2 f(r) dt^2 + \frac{1}{f(r)} dr^2 + r^2 d\Omega^2_3 \, ,
\end{equation}
where $d\Omega^2_3$ is the round metric of the $3$-dimensional sphere. The lapse function for quasitopological theories of gravity is constant on the branch that continuously connects to the Schwarzschild solution. Normalizing time at infinity sets $N=1$, so the two metric functions in the notation $g_{tt}=-h(r)$ obey $h=f$. The radial equation then integrates to \cite{Myers:2010ru}
\begin{equation}
    r^2 \left(1-f\right) + 2 \alpha \left(1-f\right)^2 - \frac{4 \lambda}{7 r^2} \left(1-f\right)^3 = \mu \, ,
\end{equation}
In these conventions $\mu=2m$ in the non-rotating limit of the Myers-Perry background.

\subsection{First-order perturbation at fixed temperature}
\label{subsect.3.1}

The NNLO action requires only the first-order metric correction, which is sourced by the Gauss--Bonnet term. Here, we follow the procedure presented in \cite{Ma:2020xwi}. Keeping the integration constant $\mu$ unchanged gives 
\begin{equation}\label{hfQTGcubic}
    h(r) = f(r) = 1-\frac{\mu}{r^2} + \frac{2 \alpha \mu^2}{r^6} + \mathcal{O}\left(\epsilon^2\right) \, .
\end{equation}
The position of the horizon and the temperature become
\begin{equation}
    \begin{split}
        r_h &= \sqrt{\mu} - \frac{\alpha}{\sqrt{\mu}} + \mathcal{O}\left(\epsilon^2\right) \, , \\
        T &= \frac{f'\left(r_h\right)}{4\pi} = \frac{1}{2\pi \sqrt{\mu}} \left(1-\frac{3\alpha}{\mu}\right) + \mathcal{O}\left(\epsilon^2\right) \, .
    \end{split}
\end{equation}
Thus, the horizon condition alone does not keep the temperature fixed. To perform the Euclidean calculation, we instead label the prescribed temperature by $T_0=(2\pi r_0)^{-1}$ and adjust the integration constant. Solving $T=T_0$ to the required order gives 
\begin{equation}\label{AdjParSSS}
    \begin{split}
        \mu &= r^2_0 - 6 \alpha + \mathcal{O}\left(\epsilon^2\right) \, , \\
        r_h &= r_0 - \frac{4\alpha}{r_0} + \mathcal{O}\left(\epsilon^2\right) \, .
    \end{split}
\end{equation}
The first relation changes the background parameter, while the second includes both this change and the horizon displacement in \Eqref{hfQTGcubic}. From this point onward, $r_0$ labels the horizon radius of the Einstein black hole with the prescribed temperature $T_0$. Since the first-order perturbation enters involves a term proportional to $\alpha$, neither the second-order metric correction nor the second-order horizon displacement is needed to obtain the action through NNLO.

\subsection{Euclidean action and thermodynamics}
\label{subsect.3.2}

Using the fixed-temperature perturbation in the NNLO formula \Eqref{IENNLOgeneric}, with the couplings included in $I_1$ and $I_2$ together with the bookkeeping parameter $\epsilon$ set to unity gives
\begin{equation}
    I^{\text{on}}_{E} = I_0\left[g_{(0)}\right] + I_1 \left[g_{(0)}\right] + I_2 \left[g_{(0)}\right] + \frac{1}{2} \delta I_1 \left[g_{(1)}\right] + \mathcal{O}\left(\epsilon^3\right) \, .
\end{equation}
Here $g_{(1)}$ includes the corrected parameter in \Eqref{AdjParSSS}. On the Schwarzschild background, the curvature insertions reduce to $\left.\chi_4\right|_{g_{(0)}} = 72 (r_0 r^{-2})^4$ and $\left.\mathcal{Z}_3\right|_{g_{(0)}} = 96 (r_0 r^{-2})^6$. Evaluating the quantities at the background and the required first-order variations, with $\text{Vol}(S^3)=2\pi^2$ gives 
\begin{equation}
    I_0\left[g_{(0)}\right] = \frac{\pi r^2_0 \, \beta}{8 G_N} \, , \, \, \,     I_1\left[g_{(0)}\right] = - \frac{9 \pi \, \alpha \beta}{4 G_N} \, , \, \, \,  I_2 \left[g_{(0)}\right] = - \frac{3 \pi \, \beta \lambda}{2 G_N r^2_0} \, , \, \, \,  \delta I_1 \left[g_{(1)}\right] = \frac{12 \pi \, \alpha^2 \beta}{G_N r^2_0} .
\end{equation}
The Einstein contribution includes, besides the vanishing bulk calculation, the Gibbons-Hawking-York (GHY) term \cite{Hawking:1975vcx,Gibbons:1976ue} and a flat reference subtraction. In the last equation, both the Gauss-Bonnet integrand and its lower integration limit are varied using \Eqref{AdjParSSS}. Including the shifted horizon is essential for this coefficient, just as it will be in the rotating calculation of the next section.

For a neutral static black hole, $I^{\text{on}}_{E} = \beta F$, where the Helmholtz free energy coincides with the Gibbs potential at zero angular velocity, yielding
\begin{equation}
    F(T_0) = \frac{\pi}{8 G_N} \left(\frac{12}{r^2_0} (4\alpha^2-\lambda) + r^2_0 - 18 \alpha \right) + \mathcal{O}\left(\epsilon^3\right) \, .
\end{equation}
Differentiating at fixed couplings, $S=-\left.(\partial F/\partial T_0)\right|_{\alpha,\lambda}$ and $M=F+T_0 S$ one gets 
\begin{equation}\label{MassEntrSSS}
    \begin{split}
        M &= \frac{3 \pi r^2_0}{8 G_N} - \frac{9 \pi \alpha}{4 G_N} - \frac{3\pi}{2G_N r^2_0} \left(4\alpha^2 - \lambda\right) + \mathcal{O}\left(\epsilon^3\right) \, , \\
        S &= \frac{\pi^2 r^3_0}{2 G_N} - \frac{6 \pi^2}{G_N r_0} \left(4 \alpha^2 - \lambda\right) + \mathcal{O}\left(\epsilon^3\right) \, .
    \end{split}
\end{equation}
There is no linear correction in $\alpha$ to the entropy at fixed temperature: the change in the horizon area cancels the explicit Gauss--Bonnet contribution to the Wald entropy. This cancellation does not imply that the entropy is unchanged when a different quantity, such as the mass, is held fixed. 

An independent check follows from the exact thermodynamic expressions in the same normalization \cite{Myers:2010ru}
\begin{equation}\label{MyersRobCQT}
    \begin{split}
        M &= \frac{3 \pi}{8 G_N} \left(r^2_h + 2 \alpha - \frac{4 \lambda}{7 r^2_h}\right) \, , \\
        T &= \frac{7 r^4_h + 4 \lambda}{2 \pi r_h \left(7 r^4_h + 28 \alpha r^2_h - 12 \lambda\right)} \, , \\
        S &= \frac{\pi^2}{2 G_N} \left(r^3_h + 12 \alpha r_h + \frac{36 \lambda}{7 r_h}\right) \, .
    \end{split}
\end{equation}
For comparison at the prescribed $T_0$, the temperature determines $r_0$
\begin{equation}
    r_h = r_0 - \frac{4 \alpha}{r_0} - \frac{16}{7 r^3_0} \left(7 \alpha^2 - \lambda\right) + \mathcal{O}\left(\epsilon^3\right) \, .
\end{equation}
Substitution into \Eqref{MyersRobCQT} reproduces \Eqref{MassEntrSSS}. This second-order horizon expansion is used only for the check, whereas the NNLO action above required the first order geometry alone. The results also coincide with the $q=0$ limit of Section \ref{sect.4}. We now apply the same prescription to rotating black holes, where the temperature and angular velocities must be fixed together and the cubic thermodynamic correction can be obtained witout constructing the second-order rotating metric.

\section{QT corrections to D=5 MP Black Hole Thermodynamics}
\label{sect.4}

In this section, we apply the NNLO method to five-dimensional rotating black holes with equal angular momenta in cubic quasitopological gravity. The Myers--Perry solution provides the background, while the first-order Gauss--Bonnet correction supplies the metric information needed to determine the thermodynamics through order $\alpha^2$ and $\lambda$.

\subsection{Theory, background and conventions}

We consider the same action as in Section \ref{sect.3}. In this setting, $[\alpha]=L^2$ and $[\lambda]=L^4$, with $\alpha=O(\epsilon)$ and $\lambda=O(\epsilon^2)$ in the perturbative counting. We retain the Einstein term and the contributions proportional to $\alpha$, $\alpha^2$ and $\lambda$. Throughout this section, $G_N$ denotes Newton's constant and $G$ the Gibbs potential. The term quasitopological identifies this covariant interaction through its special field equations in the static, spherically symmetric sector. We do not assume that the full cubic theory has second-order equations for the rotating ansatz. At the perturbative order considered here, only the Gauss--Bonnet term sources the required metric correction, and the cubic interaction is evaluated on the uncorrected background.

The two-derivative background is the five-dimensional Myers--Perry solution~\cite{Myers:1986un}
\begin{equation}\label{MPD5v1}
    \begin{split}
        ds_0^2 =& -dt^2 + \frac{2m}{\Sigma} \left(dt^2 - a \sin^2 \theta d\phi - b \cos^2\theta \, d\psi\right) + \frac{r^2 \, \Sigma}{\Delta} \, dr^2 + \Sigma \, d\theta^2 \\
        &+ \left(r^2 + a^2\right) \sin^2 \theta \, d\phi^2 + \left(r^2 + b^2\right) \cos^2\theta \, d\psi^2 \, , 
    \end{split}
\end{equation}
where 
\begin{equation}
    \Sigma \equiv r^2 + a^2 \, \cos^2 \theta + b^2 \, \sin^2 \theta \, \, \, \, , \, \, \, \, \Delta \equiv \left(r^2 + a^2\right) \left(r^2 + b^2\right) - 2 m r^2 \, . 
\end{equation}
Eq.\eqref{MPD5v1} has been written in standard Boyer-Lindquist form, with the angular range $0\leq \theta \leq \pi/2$, $0\leq \{\phi,\psi\}\leq2\pi$. The metric is parametrized by three independent integration constants $(m,a,b)$ describing the mass and two independent angular momenta, along the azimuthal angles $(\phi,\psi)$. The black hole background horizon $r_0$ is located at the largest positive root of $\Delta=0$. The corresponding thermodynamic quantities are
\begin{equation}
    \begin{split}
        &M_0 = \frac{3\pi \, m}{4 G_N} \hspace{.2cm}, \hspace{.2cm} T_0 = \frac{r^4_0 - a^2 \, b^2}{2\pi r_0 \left(r^2_0+a^2\right) \left(r^2_0+b^2\right)} \hspace{.2cm}, \hspace{.2cm} S_0 = \frac{\pi^2 \left(r^2_0+a^2\right) \left(r^2_0+b^2\right)}{2 G_N r_0} \hspace{.2 cm}, \\
        & \Omega_{a,0} = \frac{a}{r^2_0 + a^2} \hspace{.2cm}, \hspace{.2cm} \Omega_{b,0} = \frac{a}{r^2_0 + b^2} \hspace{.2cm}, \hspace{.2cm} J_{a,0} = \frac{\pi \, m \, a}{2 G_N} \hspace{.2cm}, \hspace{.2cm} J_{b,0} = \frac{\pi \, m \, b}{2 G_N} \, .
    \end{split}
\end{equation}
In the nonrotating limit, the static convention $f_0=1-\mu/r^2$ corresponds to $\mu=2m$. We now set $b=a\geq0$ and introduce the dimensionless spin parameter and radial variables
\begin{equation}
    q=\frac{a^2}{r^2_0} \, , \quad m = \frac{r^2_0 \left(1+q\right)^2}{2} \, , \quad R^2=r^2+a^2 \, , \quad \rho=\frac{R}{\sqrt{m}} \, , \quad k = \frac{2q}{1+q} \, .
\end{equation}
The nonextremal branch is $0\leq q<1$, or equivalently $0\leq k<1$, and the background outer horizon obeys $\rho_+^2=2-k$. Setting $x=\sin^2\theta$, we write the equal-spin geometry in a form where all unknown metric functions depend only on the radius \cite{Feng:2016dbw,Ma:2020xwi}
\begin{equation}\label{RotMetricAnsatz}
    \begin{split}
        ds^2 = &-\frac{H(R)}{W(R)} \, dt^2 + \frac{dR^2}{F(R)} + \frac{R^2}{4x \, (1-x)} dx^2 + R^2 x \left(1-x\right) \left(d\phi - d\psi\right)^2 \\
        &+ R^2 \, W(R) \, \left(x d\phi + \left(1-x\right) d\psi - \omega(R) \, dt\right)^2 \, .
    \end{split}
\end{equation}
The background functions are
\begin{equation}
    H_0 = F_0 = 1-\frac{2m}{R^2} + \frac{\nu^2}{R^4} \, , \quad W_0 = 1+\frac{\nu^2}{R^4} \, , \quad \omega_0 = \frac{\nu \, \sqrt{2m}}{R^4+\nu^2} \, , \quad \nu = a\sqrt{2m} \, .
\end{equation}
Thus $\nu^2/m^2 = k (2-k)$. We use $R$ in the dimensionful metric and $\rho=R/\sqrt m$ in the radial calculation, while $r_0$ labels the background horizon in Boyer--Lindquist coordinates. 

\subsection{Corrections to the background metric}

Since the cubic coupling first enters at order $\epsilon^2$, the metric correction needed for the NNLO action is sourced entirely by the Gauss--Bonnet interaction. We write
\begin{equation}\label{CoupExpan}
    \begin{split}
        H(\rho,k) &= F_0 + \frac{\alpha}{m} \, h(\rho,k) + \mathcal{O}\left(\epsilon^2\right) \, , \\
        F(\rho,k) &= F_0 + \frac{\alpha}{m} \, f(\rho,k) + \mathcal{O}\left(\epsilon^2\right) \, , \\
        W(\rho,k) &= W_0 + \frac{\alpha}{m} \, s(\rho,k) + \mathcal{O}\left(\epsilon^2\right) \, , \\
        \omega(\rho,k) &= \omega_0 + \frac{\alpha}{m^{3/2}} \, o(\rho,k) + \mathcal{O}\left(\epsilon^2\right) \, .
    \end{split}
\end{equation}
The functions $h,f,s,o$ are dimensionless. The factor $m^{-3/2}$ in the correction to $\omega$ ensures that this metric function has dimension $L^{-1}$. We solve the radial equations with $m=1$ and restore the scale through Eq.~\eqref{CoupExpan}. Writing $\delta G_{\mu\nu}$ for the coefficient of $\alpha$ in the Einstein tensor at unit $m$, the first-order equations are
\begin{equation}\label{VarGtensor}
    \delta G_{\mu\nu} + \mathcal{H}^{\text{GB}}_{\mu\nu} = 0 \, , \quad \mathcal{H}^{\text{GB}}_{\mu\nu}\left[g_{(0)}\right] = 2 R_{\mu\rho\sigma\lambda} R^{\phantom{\nu}\rho\sigma\lambda}_{\nu} - \frac{1}{2} g_{\mu\nu} R_{\rho\sigma\gamma\delta} R^{\rho\sigma\gamma\delta} \, , 
\end{equation}
where the second equality uses Ricci flatness. All curvature tensors on its right-hand side are evaluated on $g_{(0)}$. 

To solve Eq.~\eqref{VarGtensor}, we use the rational structure of the background and the factorization $F_0=(\rho^2-k)(\rho^2-2+k)/\rho^4$. We supplement finite Laurent polynomials in $\rho^2$ with $\log(1-k/\rho^2)$, which is real and regular on and outside the nonextremal outer horizon. This ansatz keeps the spin dependence and the logarithm exact. It is not a small-spin expansion or a truncated large-radius solution. Explicitly,
\begin{subequations}\label{MetricPerAnsatzRot}
    \begin{align}
        h(\rho,k) &= \sum^5_{j=-1} \mathcal{A}_j(k) \rho^{-2j} + \log\left(1-\frac{k}{\rho^2}\right) \sum^2_{j=-1} \mathcal{B}_j(k) \, \rho^{-2j} \, ,\\
        f(\rho,k) &= \sum^5_{j=-1} \mathcal{C}_j(k) \rho^{-2j} + \log\left(1-\frac{k}{\rho^2}\right) \sum^2_{j=-1} \mathcal{D}_j(k) \, \rho^{-2j} \, ,\\
        s(\rho,k) &= \sum^5_{j=-1} \mathcal{E}_j(k) \rho^{-2j} + \log\left(1-\frac{k}{\rho^2}\right) \sum^3_{j=-1} \mathcal{G}_j(k) \, \rho^{-2j} \, ,\\
        o(\rho,k) &= \frac{\sqrt{2}\nu}{\left(\rho^4+\nu^2\right)^2} \left[\sum^3_{j=-4} \mathcal{I}_j(k) \rho^{-2j} + \log\left(1-\frac{k}{\rho^2}\right) \sum^1_{j=-4} \mathcal{M}_j(k) \, \rho^{-2j}\right] \, .
    \end{align}
\end{subequations}
All coefficients, including those multiplying the logarithm, may depend on $k$. The four ansatz functions contain $11$, $11$, $12$ and $14$ coefficients, respectively. Substituting them into the $tt$, $RR$, $xx$, $t\phi$ and $\phi\psi$ equations and matching radial powers and logarithmic terms after clearing denominators gives 161 algebraic equations for these 48 coefficients. For generic $0<k<1$, the solution leaves five coefficients undetermined.

We fix the five remaining coefficients by imposing asymptotic flatness, unit normalization of the time coordinate, a non-rotating frame at infinity, and unchanged ADM mass and angular momenta. More explicitly, if $\left[X\right]_p$ denotes the coefficient of $\rho^p$ in the full large $\rho$ expansion, including the logarithm, we impose 
\begin{equation}\label{Conditions}
    \left[h\right]_2 = \left[h-s\right]_0 = \left[o\right]_0 = \left[h-s\right]_{-2} = \left[o\right]_{-4} = 0 \, .
\end{equation}
The first three conditions fix the asymptotic geometry and the time and angular coordinates, while the last two keep the ADM mass and both angular momenta equal to their background values. They do not keep the temperature or angular velocities unchanged. The resulting coefficients are listed in Table~\ref{tab:coefficients1}, and Appendix~\ref{apen.C} gives the solution before imposing these conditions.

The resulting perturbation satisfies all components of the linearized equations and obeys $h(\rho_+)=f(\rho_+)$, so $H$ and $F$ have a common horizon through first order. Its static limit is $h=f=8\rho^{-6}$, with $s=o=0$. We also repeated the calculation with enlarged bases containing 64 and 80 coefficients and recovered the same normalized functions. Appendix~\ref{app:ansatz-robustness} describes these tests and the extent of the conclusion they support.

\input{coefficient_table}
\FloatBarrier

\subsection{Horizon displacement and fixed thermodynamic sources}

Having obtained the metric corrections, we next determine their effect on the horizon and thermodynamic quantities. The asymptotic conditions imposed above keep the mass and angular momenta unchanged, whereas the horizon position, temperature, and angular velocities generally receive corrections. To apply the NNLO action formula in the grand canonical ensemble, we must adjust the background parameters so that the temperature and angular velocities remain fixed. We first compute the horizon displacement and the resulting thermodynamic variations, which will determine the required parameter shifts. The displacement of the outer horizon is obtained by expanding the horizon condition
\begin{equation}\label{PertHor}
    \rho_h = \rho_+ + \frac{\alpha}{m} \, \delta \rho_+ \, , \quad \delta\rho_+=-\frac{f(\rho_+)}{F_0^\prime(\rho_+)} \, .
\end{equation}
Here and below, primes on metric functions denote derivatives with respect to $\rho$. This displacement locates the horizon of the perturbed solution with unchanged $M,J_i$. Its temperature and angular velocities generally differ from those of the background, so keeping them fixed requires the parameter adjustments derived below. With the time coordinate normalized at infinity, the surface gravity, temperature and common angular velocity are
\begin{equation}\label{kappaTOm}
    \kappa = \frac{1}{2} \sqrt{\frac{H'\left(\rho_h\right) \, F'\left(\rho_h\right)}{m \ W\left(\rho_h\right)}} \, , \quad T = \frac{\kappa}{2\pi} \, , \quad \Omega = \omega(\rho_h) \, .
\end{equation}
To compute the temperature variation, we expand $H'$, $F'$ and $W$ at the perturbed horizon \Eqref{PertHor}. Each expansion contains both the explicit correction to the corresponding metric function and the contribution from the horizon displacement
\begin{equation}\label{ExpFpHpW}
    \begin{split}
        H'\left(\rho_h\right) =& F'_0 \left(\rho_+\right) + \frac{\alpha}{m} \left(h'\left(\rho_+\right) + \delta \rho_+ \, F''_0 \left(\rho_+\right)\right) + \mathcal{O}\left(\alpha^2\right) \, , \\
        F'\left(\rho_h\right) =& F'_0 \left(\rho_+\right) + \frac{\alpha}{m} \left(f'\left(\rho_+\right) + \delta \rho_+ \, F''_0 \left(\rho_+\right)\right) + \mathcal{O}\left(\alpha^2\right) \, , \\
        W\left(\rho_h\right) =& W_0\left(\rho_+\right) + \frac{\alpha}{m} \left(s\left(\rho_+\right)+\delta \rho_+ \, W'_0\left(\rho_+\right)\right) + \mathcal{O}\left(\alpha^2\right) \, .
    \end{split}
\end{equation}
Writing the physical variations as 
\begin{equation}\label{CorrectedT&Om}
    T = T_0 + \frac{\alpha}{m^{3/2}} \, \delta T(k) + \mathcal{O}\left(\epsilon^2\right) \, , \quad \Omega = \Omega_0 + \frac{\alpha}{m^{3/2}} \, \delta \Omega(k) + \mathcal{O}\left(\epsilon^2\right) \, , 
\end{equation}
their dimensionless coefficients are 
\begin{subequations}\label{dTdOm}
    \begin{align}
        \delta T &= \frac{T_0 \sqrt{m}}{2} \left. \left(\frac{h'+f'+2 \delta \rho_+  F''_0}{F'_0} - 
        \frac{s + \delta \rho_+ W'_0}{W_0}\right) \right|_{\rho=\rho_+} \, , \\
        \delta \Omega &= o\left(\rho_+\right) + \sqrt{m} \omega'_0 \left(\rho_+\right) \delta \rho_+ \, .
    \end{align}
\end{subequations}
Substitution gives 
\begin{subequations}\label{VarTvarOMJfix}
    \begin{align}
        \delta T (k) &= - \frac{\sqrt{2} \left(9-19k+29k^2-22k^3+6k^4\right)}{3\pi \left(2-k\right)^3\left(1-k\right)} \, , \\
        \delta \Omega (k) &= \frac{\sqrt{2k} \left(6+7k-16k^2+6k^3\right)}{3 \left(2-k\right)^{5/2} \left(1-k\right)} \, .
    \end{align}
\end{subequations}
The logarithms in the local metric corrections and in the horizon displacement cancel in these two combinations. This cancellation concerns $\delta T$ and $\delta\Omega$ for the perturbation constructed above. It does not require the other thermodynamic corrections to be free of logarithms.

The Euclidean action in the grand canonical ensemble instead compares solutions with the same temperature and angular velocities. We take these prescribed values to be the background expressions
\begin{equation}\label{Backgroundk}
    T_0 = \frac{1-q}{2\pi r_0 \left(1+q\right)} = \frac{1-k}{\pi \sqrt{2m} \left(2-k\right)} \, , \quad \Omega_0 = \frac{\sqrt{q}}{r_0 \left(1+q\right)} = \sqrt{\frac{k}{2m \left(2-k\right)}} \, .
\end{equation}
To compensate for the variations in Eq.~\eqref{VarTvarOMJfix}, we evaluate the corrected family at slightly shifted background parameters
\begin{equation}\label{NewBackground}
    \tilde{m} = m + \alpha \, \delta m(k) \, , \quad \tilde{k} = k + \frac{\alpha}{m} \delta k(k) \, . 
\end{equation}
Expanding $T_0(\tilde{m},\tilde{k})$ and $\Omega_0(\tilde{m},\tilde{k})$ gives the two linear conditions
\begin{subequations}
    \begin{align}
        \delta T - \frac{1}{2} T_0 \sqrt{m} \, \delta m + \sqrt{m} \delta k \, \partial_k T_0 &= 0 \, , \\
        \delta \Omega - \frac{1}{2} \sqrt{m} \Omega_0 \, \delta m + \sqrt{m} \delta k \, \partial_k \Omega_0 &= 0 \, .
    \end{align}
\end{subequations}
Solving these conditions gives
\begin{subequations}\label{Shiftsmk}
    \begin{align}
        \delta m &= -\frac{4 \left(9-16k+6k^2\right)}{3 \left(2-k\right)^2} \, , \\
        \delta k &= -4k - \frac{2k}{\left(1-k\right) \left(2-k\right)} \, .
    \end{align}
\end{subequations}
These parameter shifts can be included either in the metric or directly in the variation of the integrated Gauss--Bonnet insertion. We use the latter procedure, differentiating the complete background integral with its parameter-dependent lower integration limit. Once the parameters are adjusted, the horizon is still determined by $F(\rho_h)=0$ and requires no further free parameter. In the final Gibbs potential, $r_0$ and $q$ specify the prescribed $T,\Omega$ through Eq.~\eqref{Backgroundk}. 

\subsection{Evaluation of the Euclidean action}

The corrections to the metric allow us to compute the Euclidean action for cubic quasitopological gravity. We use $I^{on}_{E}=\beta G$, with the Euclidean bulk sign $-1/(16 \pi G_N)$ and $\beta=T^{-1}$. Applying \Eqref{IENNLOgeneric} with the couplings already included in $I_1$ and $I_2$ gives 
\begin{equation}\label{GenericIEQTG}
    I^{on}_E = I_0\left[g_{(0)}\right] + I_1\left[g_{(0)}\right] + I_2 \left[g_{(0)}\right] + \frac{1}{2} \delta I_1 \left[g_{(1)}\right] + \mathcal{O}(\epsilon^3) \, .
\end{equation}
Here $g_{(1)}$ includes the $\mathcal{O}(\alpha)$ perturbation at fixed sources, and the functional variation is evaluated at $g_{(0)}$, as in Section \ref{Sect:OverviewFRG}. Because the background is Ricci flat, the Einstein bulk integral vanishes. Its nonzero on-shell action comes from the GHY term with flat background substraction and matching boundary sources 
\begin{equation}\label{I0Rcubic}
    I_0\left[g_{(0)}\right] = \frac{\pi \beta \,  r^2_0 \left(1+q\right)^2}{8 G_N} \, .
\end{equation}
The higher-curvature bulk integrals converge at infinity. With the asymptotically flat falloffs and flat-reference subtraction used here, their surface contributions vanish as the boundary is taken to infinity.\footnote{In five dimensions, the orthonormal components of the Riemann tensor decay as $r^{-4}$, while the radial volume measure grows as $r^{3}$. The Gauss-Bonnet and cubic curvature radial integrals therefore decay as $r^{-5}$ and $r^{-9}$, respectively. For the Gauss-Bonnet boundary term \cite{Davis:2002gn}, the subtraction removes the leading flat-space contribution \cite{Khodabakhshi:2022zye}. The remaining surface integral and the cubic curvature boundary contributions vanish as the boundary radius tends to infinity, consistently with the asymptotic flat treatment in \cite{Reall:2019sah,Ma:2023qqj}. This does not eliminate the Einstein GHY contribution.}

On the Ricci-flat Myers-Perry background, the Gauss-Bonnet density reduces to the Kretchmann scalar. For equal angular momenta, this scalar is independent of the angular coordinates and depends only on the radius. In terms of the Boyer-Lindquist radial coordinate, it reads 
\begin{equation}\label{K0Equala}
    K_0 = \frac{96 m^2 \left(3r^4-10a^2r^2+3a^4\right)}{\left(r^2+a^2\right)^6} \, ,
\end{equation}
therefore, $I_1\left[g_{(0)}\right]$ is obtained from integrating this expression
\begin{equation}\label{I1g0Evl}
    I_{1}\left[g_{(0)}\right] = - \frac{\pi\alpha\beta}{4 G_N} \left(q^2-14q+9\right) \, .
\end{equation}
For the normalized cubic density, the vanishing six-dimensional Euler density in five dimensions implies, on the Ricci flat background, $\mathcal{Z}_3 = 1/2 R^{\phantom{\mu\nu}\rho\sigma}_{\mu\nu} R^{\phantom{\rho\sigma}\lambda\omega}_{\rho\sigma} R^{\phantom{\lambda\omega}\mu\nu}_{\lambda\omega}$. Consequently
\begin{align}\label{Z3Equala}
    \begin{split}
        \mathcal{Z}_3 \left[g_{(0)}\right] &= \frac{768 m^3 \left(r^2-a^2\right) \left(r^4-6 a^2 r^2+a^4\right)}{\left(r^2+a^2\right)^9} \, , \\
        I_2 \left[g_{(0)}\right] &= \frac{3 \pi \beta \lambda}{14 r^2_0 G_N} \frac{q^3-21q^2+35q-7}{1+q} \, .
    \end{split}
\end{align}
The remaining NNLO contribution is the first variation of the Gauss--Bonnet insertion. We first evaluate it along the perturbation constructed at fixed $M,J_i$, denoting the corresponding metric by $g_{M,J}(\alpha)$. After integrating over $\phi,\psi$ and $x$, we write 
\begin{equation}\label{I1ofU}
    I_1\left[g_{M,J}\right] = - \frac{\pi \alpha \beta}{4 G_N} U_{M,J}(\alpha) \, , \quad U_{M,J}(\alpha) = \int^{\infty}_{\rho_h} \mathcal{D}(\rho, \alpha) d\rho \, .
\end{equation}
Here $\mathcal D(\rho,\alpha)$ is the radial density obtained from $\sqrt{-g}\,\chi_4$ after removing the angular and Euclidean-time factors displayed outside $U_{M,J}$. Its background integral is $U_0=q^2-14q+9$, and $U_{M,J}$ is dimensionless and contains neither $G_N$ nor $\beta$. We keep $\beta$ outside this radial calculation and evaluate the final action at the prescribed temperature after the parameter adjustment.

For the following reduced expressions, we compute the curvature at $m=1$ and restore the expansion parameter as $\alpha/m$. The symbol $K_0$ then denotes $m^2$ times the physical scalar in Eq.~\eqref{K0Equala}, expressed in $\rho$. In these unit-$m$ coordinates, $\sqrt{-g}=\rho^3\sqrt{H/F}/2$, which gives
\begin{equation}\label{definitionD}
    \mathcal{D} = \mathcal{D}_0 + \frac{\alpha}{m} \mathcal{D}_1 + \mathcal{O}\left(\epsilon^2\right) \, , \quad \mathcal{D}_0 = \frac{\rho^3}{3} K_0 \, , \quad \mathcal{D}_1 = \frac{\rho^3}{2} \left(\delta K + \frac{h-f}{2 F_0} K_0\right) \, .
\end{equation}
The quantity $\delta K$ is the coefficient of $\alpha/m$ in the dimensionless Kretschmann scalar evaluated on $g_{M,J}(\alpha)$. Since the background Ricci tensor and scalar vanish, their perturbations begin at order $\alpha/m$ and their squares do not contribute to the first variation of $\chi_4$. Thus Eq.~\eqref{definitionD} contains all terms needed for the variation of an insertion already proportional to $\alpha$. 

Both the integrand and the lower limit vary due to the displacement of the horizon. Expanding the interval removed by the shifted horizon gives 
\begin{equation}\label{UvI}
    \begin{split}
        U_{M,J}(\alpha) &= U_0 + \frac{\alpha}{m} \int^\infty_{\rho_+} \mathcal{D}_1 \, d\rho - \int^{\rho_++(\alpha/m)\delta \rho_+}_{\rho_+} \left(\mathcal{D}_0+\frac{\alpha}{m} \mathcal{D}_1\right) d\rho + \mathcal{O}\left(\epsilon^2\right) \, \\
        &= U_0 + \frac{\alpha}{m} \left(\int^\infty_{\rho_+}\mathcal{D}_1 \, d\rho-\delta \rho_+ \, \mathcal{D}_0(\rho_+)\right) + \mathcal{O}\left(\epsilon^2\right) \, .
    \end{split}
\end{equation}
The term proportional to $\delta\rho_+$ comes from the change in the lower integration limit. It does not introduce a new boundary action at the horizon. To evaluate $\mathcal D_1$ without expanding each varied curvature contraction separately, we use the Gauss--Bonnet variational identity. For the covariant metric perturbation $\gamma_{\mu\nu}$ determined by $h,f,s,o$ at unit $m$, it reads 
\begin{equation}\label{MetDetGBVar}
    \delta \left(\sqrt{-g} \, \chi_4\right) = -\sqrt{-g} \mathcal{H}^{\mu\nu}_{\text{GB}} \, \gamma_{\mu\nu} + \partial_\mu \left(\sqrt{-g} \, \Theta^\mu\right) \, .
\end{equation}
Here $\Theta^\mu=4R^{\mu\nu\rho\sigma}\nabla_\sigma\gamma_{\nu\rho}$, and the curvature tensors, index operations and covariant derivative on the right-hand side are all evaluated on $g_{(0)}$. The correction enters through $\gamma_{\mu\nu}$. Stationarity removes the time derivative in the divergence term, while periodicity and regularity at the rotation axes make its angular integrals vanish. The remaining contribution is $\partial_\rho(\sqrt{-g}\Theta^\rho)$, so
\begin{equation}\label{D1def}
    \mathcal{D}_{1}(\rho) = b(\rho) + \frac{d v(\rho)}{d\rho} \, , \quad b(\rho) = -\frac{\rho^3}{2} \mathcal{H}^{\mu\nu}_{\text{GB}} \, \gamma_{\mu\nu} \, , \quad v(\rho)=\frac{\rho^3}{2} \, \Theta^\rho \, .
\end{equation}
Appendix~\ref{apen.D} gives the explicit functions $b(\rho)$ and $v(\rho)$ and the radial integration. Using them, Eq.~\eqref{UvI} becomes
\begin{equation}\label{UvII}
    U_{M,J}(\alpha) = U_0 + \frac{\alpha}{m} \left(\int^\infty_{\rho_+} b(\rho) \, d\rho + v(\infty)-v(\rho_+) - \delta \rho_+ \, \mathcal{D}_0(\rho_+)\right) + \mathcal{O}\left(\epsilon^2\right) \, ,
\end{equation}
with $v(\infty)=0$ for the perturbation constructed at fixed $M,J$. At the horizon, we must include both the contribution from the radial total derivative and the change in the integration region caused by the displacement of the horizon. Both effects are included when evaluating the variation of the Gauss-Bonnet insertion.

It remains to impose fixed $T$, $\Omega$. The background integral $U_0$ is independent of $m$ in five dimensions. This means that the shifts in \Eqref{Shiftsmk} add only $\delta k \partial_kU_0$ to the coefficient of $\alpha/m$ in $U_{M,J}$, giving $U_{T,\Omega}$
\begin{equation}\label{UfixedTOvI}
    U_{T,\Omega} = U_0 + \frac{\alpha}{m} \left(\int^\infty_{\rho_+} b(\rho) \, d\rho -v(\rho_+) - \delta \rho_+ \, \mathcal{D}_0(\rho_+) + \delta k \, \partial_kU_0\right) + \mathcal{O}\left(\epsilon^2\right) \, .
\end{equation}
The derivative acts on the complete background integral, including its $k$-dependent horizon. It already accounts for the change of the background curvature and the lower integration limit under the parameter redefinition. Applying the same shift again to the final result would count this effect twice. Performing the integration gives 
\begin{equation}\label{UfixedTOvII}
    U_{T,\Omega} = U_0 - \frac{8 \alpha}{\pi r^2_0} \, \mathcal{B}(q) + \mathcal{O}\left(\epsilon^2\right) \, ,
\end{equation}
where 
\begin{equation}\label{Bdefinition}
    \begin{split}
        \mathcal{B}(q) = \frac{\pi}{630 q^4 \left(1+q\right)} \big[ &-420q + 1050 q^2 +70 q^3 + 1785 q^4 -12719 q^5 + 8421 q^6 \\
        & - 1019 q^7-420 \left(1-q\right)^5 \left(1+q\right)^2 \log\left(1-q\right) \big] \, .
    \end{split}
\end{equation}
Note that the apparent poles at $q=0$ are removable 
\begin{equation}
    \mathcal{B}(q) = \pi \left(6-28 q + \frac{1801}{45} q^2 - \frac{1816}{45} q^3\right) + \mathcal{O}\left(q^4\right) \, .
\end{equation}
Multiplying $U_{T,\Omega}-U_0$ in \Eqref{UfixedTOvII} by the prefactor in \Eqref{I1ofU} and applying the NNLO factor 1/2 gives 
\begin{equation}\label{I1g1F}
    \frac{1}{2} \delta I_1 \left[g_{(1)}\right] = \frac{\beta \alpha^2}{G_N r^2_0} \mathcal{B}(q) \, .
\end{equation}
Adding up \Eqref{I0Rcubic}, \Eqref{I1g0Evl}, \Eqref{Z3Equala} and \Eqref{I1g1F}, we obtain the Euclidean on-shell action of the five-dimensional rotating black hole with equal angular momenta in cubic quasitopological gravity. Using $G=I^{on}_{E}/\beta$, the corresponding Gibbs potential up to and including order $\epsilon^2$ is 
\begin{equation}\label{GibbsPot}
    G = \frac{1}{G_N} \left(\frac{\pi r^2_0 \left(1+q\right)^2}{8} - \frac{\pi \alpha}{4} \left(q^2-14 q + 9\right) + \frac{\alpha^2}{r^2_0} \mathcal{B}(q) - \frac{3 \pi \lambda}{14 r^2_0} \frac{7-35q+21q^2-q^3}{1+q}\right) \, .
\end{equation}
For later comparison of the two NNLO contributions, we define
\begin{equation}\label{Cdefinition}
    \mathcal C(q)=-\frac{3\pi}{14}\,
    \frac{7-35q+21q^2-q^3}{1+q} \, .
\end{equation}
The spin dependence in Eq.~\eqref{GibbsPot} is exact at the retained perturbative order. Its use nevertheless requires small higher-curvature corrections, and the nonextremal expansion must be examined separately as $q$ approaches one.

\subsection{Mass, entropy and angular momentum}

On the equal angular momenta family, $J_1=J_2=J$ and
\begin{equation}
    S = -\left.\frac{\partial G}{\partial T}\right|_{\Omega} \, , \quad J = -\frac{1}{2} \left.\frac{\partial G}{\partial \Omega}\right|_{T} \, , \quad M = G + TS + 2 \Omega J \, .
\end{equation}
The factor 1/2 returns the individual angular momentum, not their sum. Derivatives are taken at fixed couplings using \Eqref{Backgroundk}. For example, $(\partial_TG)_\Omega$ is obtained by applying the two-variable chain rule in $(r_0,q)$, allowing $q$ to vary with $r_0$ so that $\Omega$ remains fixed. This gives 
\begin{align}
    \label{Mofr0q}M = &\frac{3 \pi r^2_0 \left(1+q\right)^2}{8 G_N} - \frac{\pi \alpha}{4 G_N} \left(q^2-14 q+9\right) - \frac{\alpha^2}{G_N r^2_0} \mathcal{B}(q) \nonumber\\
    &+ \frac{3 \pi \lambda}{14 G_N r^2_0} \frac{7-35q+21q^2-q^3}{1+q} + \mathcal{O}\left(\epsilon^3\right) \, , \\
    \label{Sofr0q}S = &\frac{\pi^2 r^3_0 \left(1+q\right)^2}{2 G_N} + \frac{2\pi^2 \alpha \, r_0}{G_N} q \left(7-q\right) + \frac{4\pi \alpha^2}{G_N r_0} \left(\frac{q-1}{q+1} \mathcal{B}(q) + q \mathcal{B}'(q)\right)  \nonumber\\
    &+\frac{6\pi^2\lambda}{7 G_N r_0} \frac{7 + 14 q^2 - 40q^3 + 3q^4}{\left(1+q\right)^2} + \mathcal{O}\left(\epsilon^3\right) \, , \\
    \label{Jofr0q}J = &\frac{\pi r^3_0 \sqrt{q} \, \left(1+q\right)^2}{4 G_N} - \frac{\pi \alpha r_0 \sqrt{q}}{2 G_N} \left(1-q\right) \left(7-q\right) + \frac{\alpha^2 \sqrt{q}}{G_N r_0} \left(\left(q-1\right) \mathcal{B}'(q) - \frac{4 \mathcal{B}(q)}{1+q}\right) \nonumber\\
    &+ \frac{3\pi\lambda \sqrt{q}}{7 G_N r_0} \frac{q^4-12q^3+30q^2-28q-7}{\left(1+q\right)^2} + \mathcal{O}\left(\epsilon^3\right) \, .
\end{align}
In \Eqref{Mofr0q} - \Eqref{Jofr0q}, primes denote derivatives with respect to $q$ and all quantities are expressed in terms of the same parameters $r_0$, $q$ used in the Gibbs potential\footnote{The parameters specify the temperature and angular velocities through the background relations, so $r_0$ should not be identified with the corrected horizon radius. Since both $T$ and $\Omega$ scale as $r^{-1}_0$ at fixed $q$, the thermodynamic relations imply $M=G+r_0 \partial_{r_0} \left.G\right|_{q}$, providing a useful check of the mass expression.}. To compare entropies at fixed ADM charges, we adjust the parameters of the corrected solution such that its mass and individual angular momenta match those of a chosen reference Myers-Perry black hole. Denoting the parameters of this reference solution by $(\bar{r}_0,\bar{q})$, we obtain the difference of entropies at the same mass $M_0(\bar{r}_0,\bar{q})$ and individual angular momenta $J_0(\bar{r}_0,\bar{q})$
\begin{equation}
    \begin{split}
        \left.\Delta S\right|_{M,J_i} = &\frac{\pi^2 \alpha \bar{r}_0 \left(1+\bar{q}\right) \left(\bar{q}^2-14\bar{q}+9\right)}{2 \left(1-\bar{q}\right) G_N} + \frac{3 \pi^2 \lambda \left(7-35 \bar{q}+21\bar{q}^2-\bar{q}^3\right)}{7 \bar{r}_0 \left(1-\bar{q}\right) G_N} \\
        &+\frac{\pi^2 \alpha^2}{12 G_N \bar{r}_0 \left(\bar{q}-1\right)^3 \left(\bar{q}+1\right)} \big[5\bar{q}^6 + 12 \bar{q}^5 - 211 \bar{q}^4 + 672 \bar{q}^3 - 769\bar{q}^2 + 564\bar{q} \\
        &-81 + 24 \pi^{-1} \left(1-\bar{q}^2\right)^2 \mathcal{B}(\bar{q}) \big] + \mathcal{O}\left(\epsilon^3\right) \, .
    \end{split}
\end{equation}
The NNLO fixed charge term includes the quadratic effect of the first-order parameter redefinition. It cannot be inferred by replacing $\Delta S$ by $-\Delta G/T_0$ beyond first order.

\subsection{Thermodynamic properties and consistency checks}

We have verified that the metric corrections satisfy the first-order radial field equations and checked the radial integration by differentiating the analytical result before evaluating its limits. As an independent thermodynamic check, we calculated the Wald entropy from the corrected horizon geometry, including its displacement. At fixed mass and angular momenta, this agrees with the entropy derived from the Gibbs potential through first order in $\alpha$, as shown in Appendix~\ref{app:wald-first-order}. 

The static limit provides a further check of the thermodynamic results. Setting $q=0$ in Eqs.~\eqref{Mofr0q} and~\eqref{Sofr0q}, we obtain 
\begin{equation}
\begin{split}
        \left.M\right|_{q=0}&=\frac{3\pi r_0^2}{8G_N}-\frac{9\pi\alpha}{4G_N}+\frac{3\pi\lambda-12\pi\alpha^2} {2 G_N r_0^2}+O(\epsilon^3) \, , \\
        \left.S\right|_{q=0}&=\frac{\pi^2r_0^3}{2G_N} +\frac{6\pi^2\lambda-24\pi^2\alpha^2}{G_N \, r_0}+\mathcal{O}(\epsilon^3) \, .
\end{split}
\end{equation}
%
For $\lambda=0$, these expressions reproduce the expansion of the exact static Einstein--Gauss--Bonnet solution \cite{Boulware:1985wk} at fixed
$T=(2\pi r_0)^{-1}$.\footnote{In the normalization of Eq.~\eqref{RotatingAction}, the exact static solution has $T=R_h/[2\pi(R_h^2+4\alpha)]$, $M=3\pi(R_h^2+2\alpha)/(8G_N)$ and $S=\pi^2(R_h^3+12\alpha R_h)/(2G_N)$. Fixing the temperature gives $R_h=(r_0+\sqrt{r_0^2-16\alpha})/2$ on the branch connected to Schwarzschild. Expanding this root in the exact expressions reproduces the terms through $\alpha^2$ above. Here $R_h$ is the static horizon radius, which coincides with the areal radius.} This gives an independent check of the Gauss--Bonnet contributions without requiring a numerical extrapolation of the rotating solution.

\begin{figure}[t!]
    \centering
    \includegraphics[width=0.9\linewidth]{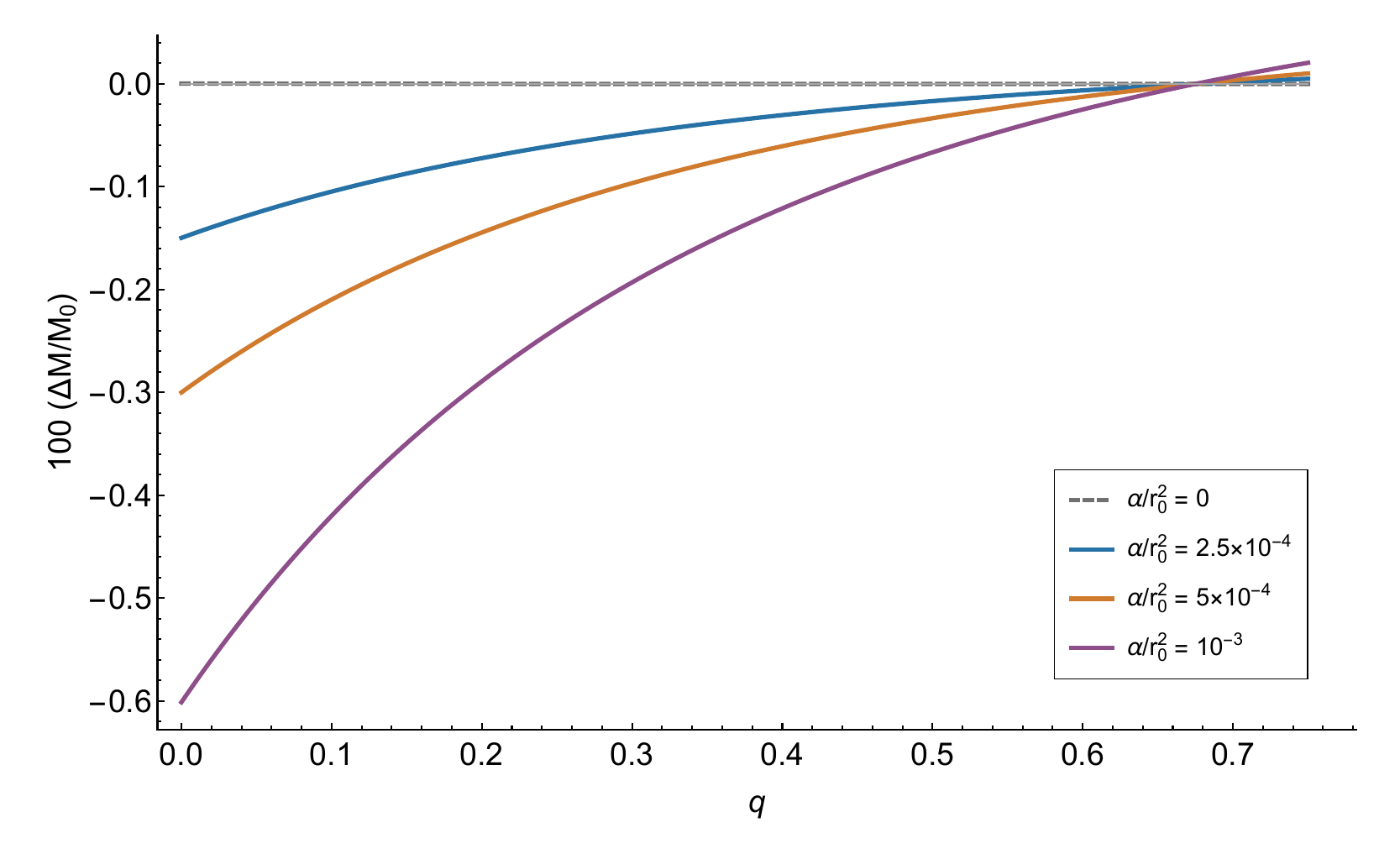}
    \caption{
    Relative mass correction $100(M-M_0)/M_0$ for solutions compared at the same $T$ and $\Omega$, as a function of $q=a^2/r_0^2$. The curves use $\alpha/r_0^2=0,\,2.5\times10^{-4},\,5\times10^{-4},\,10^{-3}$ and $\lambda/r_0^4=10^{-6}$, retaining terms of order $\alpha$, $\alpha^2$ and $\lambda$. Here $M_0$ is the Einstein mass at the same sources. The dashed curve retains the cubic interaction, while the horizontal zero line represents the Einstein reference. The scale $r_0$ is constant along the scan, so $T$ and $\Omega$ vary with $q$ even though they coincide between the solutions compared at each point.}
    \label{fig:MassDiff}
\end{figure}

\begin{figure}[t!]
    \centering
    \includegraphics[width=0.9\linewidth]{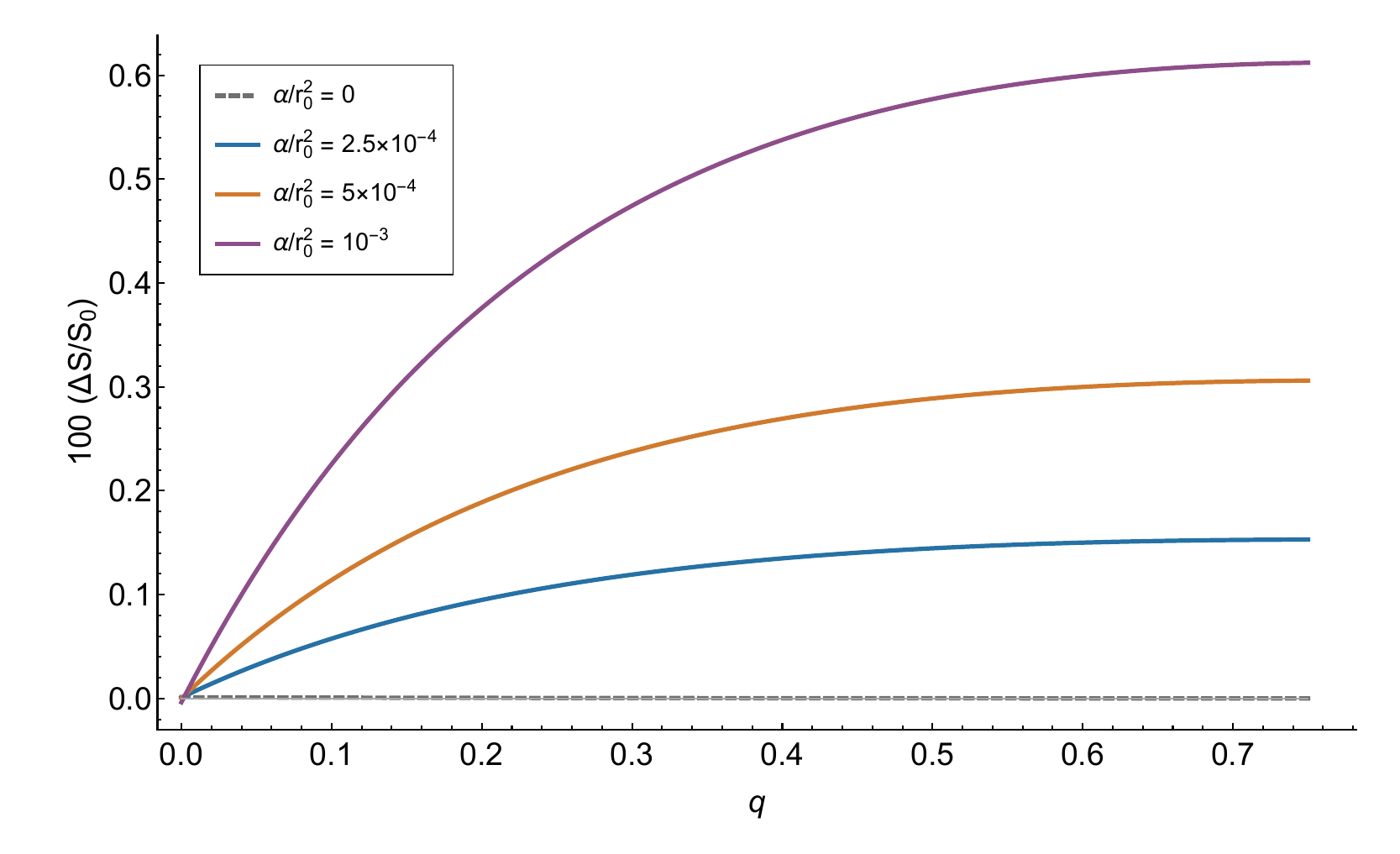}
    \caption{Relative entropy correction at the same $T$ and $\Omega$, using the couplings and comparison prescribed in Fig.~\ref{fig:MassDiff}. Here $S_0$ denotes the Einstein entropy at those sources. The linear Gauss--Bonnet contribution vanishes in the static limit, while the $\alpha^2$ and $\lambda$ terms produce the small, coupling-dependent values at $q=0$. The dashed $\alpha=0$ curve contains only the cubic correction at the retained order, and the horizontal zero line is the Einstein reference.}
    \label{fig:EntroDiff}
\end{figure}

We can now use the Gibbs potential~\eqref{GibbsPot} to examine how rotation affects the size and sign of the thermodynamic corrections. Figures~\ref{fig:MassDiff} and~\ref{fig:EntroDiff} compare the corrected mass and entropy with their Einstein values at the same temperature and angular velocity. Each curve corresponds to a different Gauss--Bonnet coupling, while the cubic coupling is fixed. The sources $T$, $\Omega$ coincide between the two solutions at each value of $q$, although their values change as we move along the horizontal axis with $r_0$ held constant.

For positive $\alpha$, the leading Gauss--Bonnet correction lowers the mass at small $q$, and the size of this decrease grows with $\alpha$. As rotation increases, the mass difference becomes less negative and the leading term changes sign at $q=7-2\sqrt{10}\simeq0.68$. For the nonzero values of $\alpha$ shown in Fig.~\ref{fig:MassDiff}, the NNLO terms shift this crossing only slightly. The $\alpha=0$ curve instead isolates the cubic contribution and has its zero at $q\simeq0.23$. These sign changes indicate whether the corrected solution has a larger or smaller mass than the Einstein solution with the same sources, rather than a change in the sign of the total mass.

The leading entropy correction behaves differently: for positive $\alpha$ it is positive throughout $0<q<1$, and its fractional contribution grows over the interval shown in
Fig.~\ref{fig:EntroDiff}. At $q=0$, the change in horizon area
cancels the explicit Gauss--Bonnet contribution to the Wald entropy at first order, explaining the absence of a linear term in  $\alpha$ in the static expression above.
The NNLO terms, nevertheless, leave a static correction whose sign depends on the relative values of $\alpha^2$ and $\lambda$. Thus the vanishing static correction applies only at first order, and the full entropy difference need not be positive at small spin.

\begin{figure}[t!]
    \centering
    \includegraphics[width=0.9\linewidth]{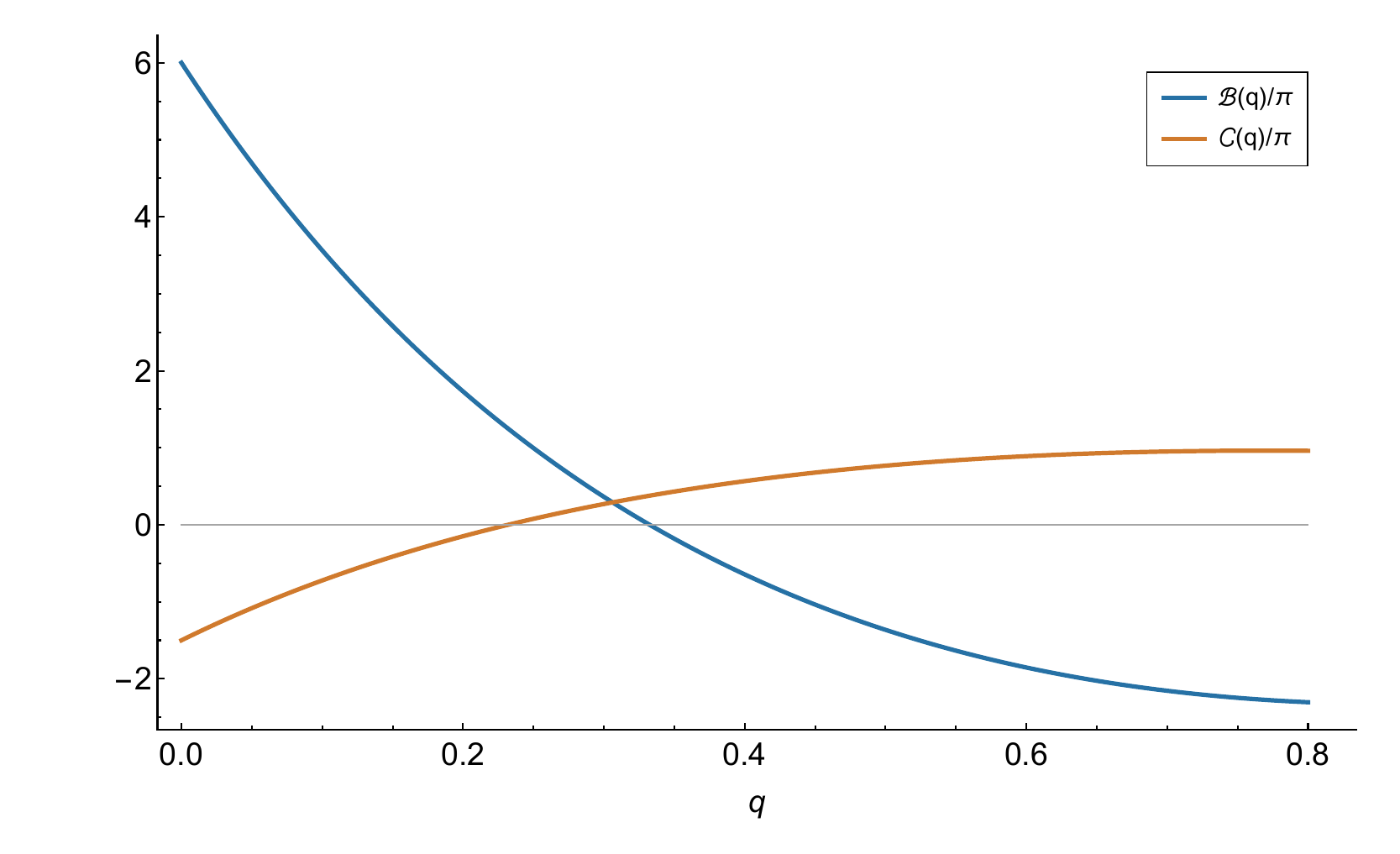}
    \caption{Spin dependence of $\mathcal{B}(q)/\pi$ and $\mathcal{C}(q)/\pi$, the coefficients of the quadratic Gauss--Bonnet and linear cubic contributions to the Gibbs potential, respectively. Their normalization is defined by $\Delta G_{\mathrm{NNLO}} =[\alpha^2\mathcal{B}(q)+\lambda\mathcal{C}(q)]/(G_Nr_0^2)$. No coupling values are selected in this figure, and the full logarithmic dependence of $\mathcal{B}(q)$ is included.}
    \label{fig:BCGibbs}
\end{figure}

To distinguish the two NNLO contributions, Fig.~\ref{fig:BCGibbs} shows the coefficients $\mathcal{B}(q)$ and $\mathcal{C}(q)$ that multiply $\alpha^2/(G_Nr_0^2)$ and $\lambda/(G_Nr_0^2)$, respectively, in the Gibbs potential. The Gauss--Bonnet coefficient starts at $\mathcal{B}(0)=6\pi$ and changes sign at $q\simeq0.33$, whereas the cubic coefficient starts at $\mathcal{C}(0)=-3\pi/2$ and changes sign at $q\simeq0.23$. Their spin dependence therefore changes the relative contribution of the two interactions to the free energy. Since the entropy follows from a derivative of the potential at fixed $\Omega$, these curves are not the corresponding entropy coefficients. In particular, their zeros neither imply vanishing total entropy nor establish a thermodynamic phase transition.

%
%

The logarithmic correction deserves a separate discussion. The part of the entropy proportional to $\log(1-q)$ in the comparison at fixed $T,\Omega$ is
\begin{equation}
    \left.S_{\text{log}}\right|_{T,\Omega} = \frac{8 \pi^2 \alpha^2}{3 r_0 G_N} \frac{\left(1-q\right)^4 \left(5+2q+3q^2\right)}{q^4} \log\left(1-q\right) \, .
\end{equation}
This term arises from the classical radial solution and its action integral and should not be confused with logarithmic quantum corrections to black-hole entropy~\cite{Sen:2012dw}.
Although the logarithmic and rational terms are separately singular at $q=0$, these singularities cancel in their sum, leaving a real entropy for $0\leq q<1$ with a regular static limit. The two contributions must therefore be combined before taking that limit. 

Near $q=1$, the logarithmic contribution vanishes as $(1-q)^4\log(1-q)$, so this term does not produce an entropy divergence. A finite limit of the thermodynamic expressions, however, is not enough to establish the validity of the perturbative expansion: the shifts used to maintain the prescribed $T,\Omega$ become singular there. The nonextremal results therefore motivate a separate calculation of the extremal near-horizon geometry, with both entropies expressed in terms of the same conserved angular momentum before comparing them.

\section{Conclusions}
\label{sect.Concl}

In this paper, we have computed higher-curvature corrections up to (and including) cubic order in curvature of five-dimensional, asymptotically flat black holes with equal angular momenta through NNLO. The application of the Reall-Santos method \cite{Reall:2019sah} allows to obtain, for equal angular momenta, analytic expressions for the thermodynamic potential that greatly simplify the corresponding thermodynamic quantities in the grand canonical ensemble. For Einstein gravity supplemented by Gauss--Bonnet and cubic quasitopological interactions, the Gibbs potential includes the contributions proportional to the couplings $\alpha$, $\alpha^2$ and $\lambda$, retaining the full dependence on the rotation parameter without resorting to a slowly rotating approximation. Its thermodynamic derivatives determine the corrected mass, entropy and angular momenta. 

The calculation illustrates how the perturbative method for the Euclidean action reduces the geometric information needed to obtain these quantities. At NNLO, the first-order correction to the metric induced by the Gauss--Bonnet term is sufficient, while the cubic interaction enters through its evaluation on the Myers-Perry background. Implementing this prescription requires fixing certain thermodynamic quantities determined by the thermodynamic ensemble. In this work, the ensemble corresponds to the grand canonical ensemble, which fixes the temperature and angular velocity. We incorporate this prescription by adjusting the background parameters of the Einstein solution. The static calculation provides a direct check against the exact thermodynamic expressions obtained in \cite{Myers:2010ru} and agrees with the nonrotating limit of the Gibbs potential. 

Although we have used the standard cubic quasitopological interaction to connect the static and rotating calculations, the NNLO thermodynamic results apply more broadly. On a five dimensional Ricci-flat background, the two purely Riemann-cubic invariants satisfy a dimensional identity, so arbitrary algebraic cubic curvature interactions contribute through the single combination $\lambda_7 + 2\lambda_8$ defined in \Eqref{Z3Rot}. The results therefore retain their structure with this combination replacing $\lambda$. The same holds for arbitrary quadratic interactions, which differ from a multiple of the Gauss--Bonnet density only by the squares of the Ricci tensor and the Ricci scalar. Since these terms vanish on the Ricci-flat background together with their first variation, they neither source the first-order correction to the metric nor contribute through NNLO, and $\alpha$ is simply the coefficient of the Riemann-squared term. This statement concerns the thermodynamics at the retained order and does not require the rotating field equations to share the simplifications of the static, spherically symmetric part. 

The resulting corrections depend on both the rotation and the thermodynamic quantities held fixed in the comparison. In particular, the entropy correction linear in $\alpha$ vanishes at fixed temperature in the static limit, whereas rotation produces a nonzero contribution. The logarithms appearing at the NNLO originate from the classical metric perturbations, and their apparent singularities in the static limit $q \rightarrow 0$ cancel out in the complete expressions. Near extremality, however, finite limits of individual thermodynamic coefficients are not enough to establish the validity of the perturbative expansion, since the parameter shifts that keep the temperature and angular velocity fixed become singular in that limit. A reliable description of this regime requires a separate analysis of the extremal near-horizon geometry, in which the corrected and uncorrected entropies are compared at the same angular momentum~\cite{Ma:2020xwi,Horowitz:2023xyl,Mao:2023qxq,Cano:2023dyg}.

Our results suggest several directions for future work. The first one is a more detailed study of the free energy. The closed expressions obtained here give access to the response functions of the black hole at fixed temperature and angular velocity, and therefore to its local thermodynamic stability, and they can be translated to the ensembles in which the angular momenta are fixed. As emphasized above, the sign changes found in the individual corrections do not by themselves establish a phase transition. It would be interesting to determine whether the higher-curvature terms modify the stability properties~\cite{Monteiro:2009tc} and the phase structure of rotating black holes, which is known to be nontrivial already for asymptotically flat Myers--Perry solutions~\cite{Altamirano:2014tva}, and how the answer depends on the ensemble.

A second direction is the extension to unequal angular momenta. In that case the metric functions no longer depend only on the radius, and the first-order correction to the metric is only known as an expansion in the spins~\cite{Ma:2025vjk}, while the leading correction to the thermodynamics is available for general rotation parameters~\cite{Wu:2024iiz}. Obtaining the NNLO potential in this setting would show which of the features found here, such as the sign changes of the corrections or the classical logarithms, are specific to equal angular momenta. Finally, Myers--Perry black holes with all angular momenta equal are cohomogeneity-one in every odd dimension~\cite{Kunduri:2006qa}, that is, their metric functions depend only on the radial coordinate, so the strategy followed here is not restricted to $D=5$. It would be interesting to repeat the calculation in higher odd dimensions and to identify how the corrections depend on $D$, with the aim of expressing the NNLO thermodynamic quantities in closed form for general dimension. We leave these questions for future work.

\acknowledgments
\label{sect.Conclu}
I would like to thank Pablo Bueno and \'Angel Rinc\'on for useful discussions, and David Kubiz\v{n}\'ak for suggesting a closer look at the free energy and the phase structure of these solutions. I would also like to thank the Institute of Mathematics of the Czech Academy of Sciences, where part of this work was carried out, for its hospitality, and its members for stimulating discussions and comments. This work is supported by ANID through fellowship No.~74250068.


\appendix

\section{First-order metric correction}\label{apen.C}

\subsection{General perturbative solution}

Within the ansatz of Eq.~\eqref{MetricPerAnsatzRot}, the first-order field equations leave five coefficients free, denoted by $\mathcal A_0$, $\mathcal A_1$, $\mathcal E_0$, $\mathcal E_2$ and $\mathcal I_{-2}$. The functions before imposing the asymptotic conditions are

\begin{equation}
    \begin{split}
        h(\rho,k) =& \frac{k\left(k-2\right)}{3\rho^4} \left[\frac{3 \, \mathcal{E}_2}{k \left(k-2\right)} - 3 \mathcal{A}_0 + 3 \mathcal{E}_0 - \frac{32 \left(k^2-8k-2\right)}{k^2 \left(k-2\right)^2}\right] + \frac{\mathcal{A}_1}{\rho^2} + \mathcal{A}_0 \\
        &-\frac{16 k \left(k-2\right)}{3 \rho^8} + \frac{8 \left(2k^2-4k+7\right)}{3 \rho^6} - \rho^2 \left(\mathcal{E}_0 + \frac{32}{k \left(k-2\right)^2}\right) \\
        &-\frac{32}{k \left(k-2\right) \rho^4} \left[1 - \frac{\rho^2 \left(5k^2 - 10k -4\right)}{3 k \left(k-2\right)} + \frac{\rho^4 \left(2k^2-4k-7\right)}{3k \left(k-2\right)}\right] \log\left(1-\frac{k}{\rho^2}\right) \\
        &-\frac{32 \rho^2}{k^2 \left(k-2\right)^2} \log\left(1-\frac{k}{\rho^2}\right) \, , 
\end{split}
\end{equation}

\begin{equation}
\begin{split}
        f(\rho,k) =& - \frac{32 k^2 \left(k^2-4k+4\right)}{3 \rho^{10}} - \frac{16 k \left(k-2\right)}{\rho^8} + \frac{8 \left(2k^2-4k+1\right)}{\rho^6} + \frac{1}{\rho^4} \left(\mathcal{E}_2 - \frac{64}{3k \left(k-2\right)}\right) \\
        & +\frac{1}{\rho^2} \left[\mathcal{A}_1 + 2 \mathcal{A}_0 - 2 \mathcal{E}_0 \left(k^2-2k+1\right) - \frac{32 \left(4k^2-11k+16\right)}{3 k \left(k-2\right)^2}\right] - \frac{64}{k \left(k-2\right)^2} \\
        & - 3\mathcal{E}_0 + \rho^2 \left(\mathcal{E}_0 + \frac{32}{k \left(k-2\right)^2}\right) - \frac{32 \left(2k^2-4k-1\right)}{3 k \left(k-2\right) \rho^4} \log \left(1-\frac{k}{\rho^2}\right) \\
        & + \frac{32 \rho^2}{k^2 \left(k-2\right)^2} \left[1 - \frac{5k^2-10k-4}{3\rho^4} - \frac{3}{\rho^2}\right] \log \left(1-\frac{k}{\rho^2}\right) 
\end{split}
\end{equation}

\begin{equation}
\begin{split}
        s(\rho,k) =& \frac{16 k^2 \left(k^2-4k+4\right)}{3 \rho^{10}} + \frac{16 k \left(k-2\right)}{3 \rho^8} + \frac{16 \left(3k^2-6k-1\right)}{3\rho^6} + \frac{\mathcal{E}_2}{\rho^4} + \mathcal{E}_0 \\
        & - \frac{k \left(k-2\right)}{\rho^2} \left(\mathcal{E}_0 + \frac{16 \left(2k+1\right)}{k^2 \left(k-2\right)^2}\right) - \rho^2 \left(\mathcal{E}_0 + \frac{32}{k \left(k-2\right)^2}\right) \\
        &+\frac{32}{k \left(k-2\right)} \left(\frac{1}{k \left(k-2\right)}-\frac{1}{\rho^2}-\frac{\rho^2}{k \left(k-2\right)}-\frac{2k^2-4k-1}{3 \rho^4}\right) \log \left(1-\frac{k}{\rho^2}\right) \, . 
\end{split}
\end{equation}

\begin{equation}
\begin{split}
        o(\rho,k) =& \frac{\sqrt{2 k \left(2-k\right)} \rho^8}{k \left(2-k\right) \left(k \left(2-k\right) + \rho^4\right)^2} \Bigg( \frac{\left(k^2-2k+6\right) \mathcal{E}_0}{8} + \frac{\mathcal{E}_2}{4 k \left(k-2\right)} + \frac{\mathcal{A}_1}{8} + \frac{\mathcal{I}_{-2}}{2} \\
        &- \frac{6k^2-24k-11}{3k \left(k-2\right)} \Bigg) + \frac{\sqrt{2k \left(2-k\right)}}{\left(\rho^4-k \left(k-2\right)\right)^2} \Bigg[ \frac{k \left(k-2\right) \mathcal{A}_1}{8} - \frac{k \left(k-2\right)}{2} \mathcal{I}_{-2} \\
        &+ \frac{k \left(k^3-4 k^2+2k+4\right) \mathcal{E}_0}{8} + \rho^4 \mathcal{I}_{-2} + 2 \rho^6 \left(\mathcal{E}_0+\frac{24}{k \left(k-2\right)^2}\right) - \frac{3}{4} \mathcal{E}_2 \\
        &-\frac{64 k \left(k-2\right)}{3 \rho^2} - \frac{6k^2-24k+5}{3} + \frac{16 \left(3k+4\right) \rho^2}{3 k \left(k-2\right)} + \Bigg( 16 - \frac{16 \rho^8}{k^2 \left(k-2\right)^2} \\
        &+ \frac{64 \left(k^2-2k-2\right) \rho^4}{3 k^2 \left(k-2\right)^2} + \frac{64 \rho^6}{k^2 \left(k-2\right)^2} \Bigg) \log \left(1-\frac{k}{\rho^2}\right) \Bigg] \, .
    \end{split}
\end{equation}

The conditions in Eq.~\eqref{Conditions} fix these five coefficients simultaneously, giving
\begin{align}
 \mathcal A_0=\mathcal E_0&=-\frac{32}{k(k-2)^2},&
 \mathcal A_1&=-\frac{16(4k^2-5k-14)}{3k(k-2)^2},\nonumber\\
 \mathcal E_2&=\frac{4(6k^3-16k^2+15k-30)}{3(k-2)},&
 \mathcal I_{-2}&=-\frac{8(k-8)}{k(k-2)^2}.
 \label{eq:C-free-values}
\end{align}
Substituting these values recovers the coefficients in Table~\ref{tab:coefficients1}. The apparently singular powers of $k$ cancel only after the rational and logarithmic pieces are combined, so the static limit must be taken in the complete functions. For $0<k<1$, the logarithm is regular throughout the exterior $\rho^2\geq2-k$. These expressions give the general solution within the chosen rational and logarithmic ansatz, without asserting that this ansatz exhausts every solution of the radial equations.

\subsection{Robustness under enlargement of the ansatz}
\label{app:ansatz-robustness}

To test the dependence of the solution on the chosen summation
bounds, we enlarge all eight sums in
Eq.~\eqref{MetricPerAnsatzRot}. Denoting their original bounds
by $j_{\min}$ and $j_{\max}$, we replace them by
$j_{\min}-p$ and $j_{\max}+n$, respectively, where $p,n$
are nonnegative integers. The enlarged basis contains
$48+8(p+n)$ coefficients. Increasing $n$ adds higher inverse
powers of $\rho$, whereas increasing $p$ allows additional
growing powers before imposing the boundary conditions.

We tested $(p,n)=(0,2),(0,4),(2,2)$, corresponding to bases
containing 64, 80, and 80 coefficients, respectively.
All coefficients, including those in the original ansatz,
were solved again from the linearized equations, keeping
the logarithm exact. In every case, five coefficients remain
undetermined before asymptotic normalization and are
completely fixed by the boundary conditions. All newly
introduced coefficients vanish in the normalized solution,
and the four functions agree identically with the original
result:
\begin{equation}
 (h,f,s,o)_{p,n}-(h,f,s,o)_{0,0}=(0,0,0,0).
\end{equation}
Consequently, these enlargements do not modify the
fixed-source perturbation or its contribution to the
NNLO Gibbs free energy.

The resulting functions satisfy the full linearized radial equations, so the finite ansatz is not a truncated large-radius solution. Nevertheless, these tests
establish stability only within the finite spaces examined.
They neither prove uniqueness among all possible radial
functions nor establish a limit as the summation bounds
tend to infinity.
\newpage
\section{Radial integration of the Gauss--Bonnet variation}\label{apen.D}

This appendix gives the functions in Eq.~\eqref{D1def} for the perturbation constructed at fixed $M,J_i$. All curvature tensors in $\mathcal H_{\mathrm{GB}}^{\mu\nu}$ and $\Theta^\rho$ are evaluated on the background metric. The first-order correction enters through $\gamma_{\mu\nu}$, and the density factor is already included in $b$ and $v$. We use $m=1$ and $0<k<1$ throughout. The bulk contribution is

\begin{equation}
    b(\rho) = \frac{256}{3 \left(2-k\right)^2 \, k^2 \, \rho^{15}\left(k-\rho^2\right)} \left[\mathcal{P}_{b}(\rho)+\log\left(1-\frac{k}{\rho^2}\right) \, \mathcal{Q}_{b}(\rho)\right] \, ,
\end{equation}

with $\mathcal{P}_{b}$ and $\mathcal{Q}_{b}$ given by 

\begin{equation}
    \begin{split}
        \mathcal{P}_{b}(\rho) =& 6 k \rho^{12} + 3 k \, \rho^{10} \left(22k^2-45k-4\right) + \left(2-k\right) \left(6k^3-88k^2+192k+75\right) \, k^2 \, \rho^8 \\
        &-2 k^3 \, \rho^6 \left(2-k\right)^2 \left(9k^3-39 k^2+74k+24\right) - 2k^4 \, \rho^4 \left(2-k\right)^3 \left(9k^2-33k+26\right) \\
        &+6 \left(7-k\right) \left(2-k\right)^4 \, k^5 \, \rho^2 - 6 \left(2-k\right)^5 \, k^6 \, .
\end{split}
\end{equation}

\begin{equation}
\begin{split}
        \mathcal{Q}_{b}(\rho) = & 6 \rho^{14} + 6 \rho^{12} \left(11k^2-23k-2\right) + 6k \, \rho^{10} \left(12k^3 - 59 k^2 + 58 k + 26\right) \\
        &-24 k^2 \, \rho^8 \left(2-k\right) \left(k^3-7k^2+8k+7\right) - 24 k^3 \, \rho^6 \left(2-k\right)^2 \left(k^2-2k-2\right) \, .
    \end{split}
\end{equation}

The radial total derivative in Eq.~\eqref{D1def} is determined by

\begin{equation}
    v(\rho) = \frac{32}{3 \left(2-k\right)^2 k^2 \, \rho^{14}} \left[\mathcal{P}_{v}(\rho)+\log\left(1-\frac{k}{\rho^2}\right) \, \mathcal{Q}_{v}(\rho)\right]
\end{equation}

with $\mathcal{P}_{v}$ and $\mathcal{Q}_{v}$ given by 

\begin{equation}
    \begin{split}
        \mathcal{P}_v(\rho) =& -96 k \, \rho^{10} + k^2 \, \rho^6 \left(2-k\right)^2 \left(36k^4-96k^3+114k^2-228k-5\right) \\
        &+ 48 k \, \rho^8 \left(2-k\right) \left(1+5k\right) - 4 k^3 \, \rho^4 \left(2-k\right)^3 \left(18k^2-36k+41\right) \\
        &- 24 k^5 \left(2-k\right)^5 + 122k^4 \, \rho^2 \, \left(2-k\right)^4 
\end{split}
\end{equation}

\begin{equation}
\begin{split}
        \mathcal{Q}_v(\rho) =& 32 k \, \rho^8 \left(2-k\right) \left(7k^2-14k-5\right) - 48 \rho^{10} \left(5k^2-10k-2\right) \\
        &-96 \rho^{12} - 48 k^2 \, \rho^6 \left(2-k\right)^2 \left(2k^2-4k-1\right) \, .
    \end{split}
\end{equation}

Expanding the complete expression at large $\rho$ gives $v(\infty)=0$. At the background horizon, $\rho_+^2=2-k$, it gives 

\begin{equation}
    \begin{split}
        v(\rho_+) = \frac{32}{3 (2-k)^4 \, k^2} \Bigg[ &k\left(36k^5-192k^4+380k^3-392k^2+235k-48\right) \\
        &-32\left(1-k\right)^2 \left(3k^2-7k+3\right) \log\left(\frac{2(1-k)}{2-k}\right)\Bigg] \, .
    \end{split}
\end{equation}

To integrate $b(\rho)$, set $u=\rho^{-2}$. A function whose derivative equals the transformed integrand is
\begin{equation}\label{eq:D-integrated-function}
 \mathcal P(u,k)=\frac{128}{3k^2(2-k)^2}
 \left[\sum_{n=1}^{7}a_n(k)u^n+
 \log(1-ku)\sum_{n=0}^{4}b_n(k)u^n\right],
\end{equation}
where these auxiliary coefficients are used only to present the integral compactly:
\begin{align}
 a_1&=\frac{4-16k+23k^2-20k^3+6k^4}{k},\nonumber\\
 a_2&=\frac{4-28k+23k^2-20k^3+6k^4}{2},\nonumber\\
 a_3&=\frac{k}{3}(22+173k-76k^2-20k^3+6k^4),\nonumber\\
 a_4&=(2-k)^2k^2(-11-24k+3k^2),\nonumber\\
 a_5&=\frac{2}{5}(k-2)^3k^3(-4-18k+9k^2),\nonumber\\
 a_6&=5(2-k)^4k^4,\qquad
 a_7=\frac{6}{7}(k-2)^5k^5,
\end{align}
and
\begin{align}
 b_0&=\frac{4-16k+23k^2-20k^3+6k^4}{k^2},&
 b_1&=-6,\nonumber\\
 b_2&=6+66k-33k^2,&
 b_3&=-24k(2+3k-4k^2+k^3),\nonumber\\
 b_4&=-6(2-k)^2k^2(-2-2k+k^2).
\end{align}
These $b_n(k)$ are coefficients of $\mathcal P$ and are distinct from the radial function $b(\rho)$. Direct differentiation verifies
\begin{equation}
 \frac{\partial\mathcal P}{\partial u}
 =\frac{b(u^{-1/2})}{2u^{3/2}},\qquad
 \mathcal P(0,k)=0.
\end{equation}
The reversal of the integration limits compensates the minus sign in $d\rho=-du/(2u^{3/2})$, giving
\begin{equation}
 \int_{\rho_+}^{\infty}b(\rho)\,d\rho
 =\int_0^{1/(2-k)}\frac{b(u^{-1/2})}{2u^{3/2}}\,du
 =\mathcal P\left(\frac{1}{2-k},k\right).
\end{equation}
Combining this result with the total derivative and the displaced lower limit yields
\begin{equation}
 U_{M,J}=U_0+\frac{\alpha}{m}
 \left[
 \mathcal P\left(\frac{1}{2-k},k\right)
 -v(\rho_+)-\delta\rho_+\mathcal D_0(\rho_+)
 \right]+O(\epsilon^2) \, .
\end{equation}
Finally, adding $(\alpha/m)\delta k\,\partial_kU_0$ as in Eq.~\eqref{UfixedTOvI} gives the variation at fixed $T,\Omega$. With $k=2q/(1+q)$, the horizon logarithm becomes $\log(1-q)$, and the result reduces to Eqs.~\eqref{UfixedTOvII} and~\eqref{Bdefinition}.

\newpage


\section{Wald entropy at first order}
\label{app:wald-first-order}

We evaluate the entropy on the perturbative solution constructed
with unchanged ADM mass and angular momenta. The parameters
$\bar r_0$ and $\bar q$ label the reference Myers--Perry black hole
with the same charges. Through first order, the Wald entropy is \cite{Wald:1993nt,Iyer:1994ys}
\begin{equation}
    S_{\mathrm{Wald}}
    =
    \frac{A_{\mathcal H}}{4G_N}
    + \frac{\alpha}{2G_N}
    \int_{\mathcal H} d^3x\,\sqrt{\sigma}\,
    \mathcal R_{\mathcal H}
    + \mathcal O(\epsilon^2),
    \label{eq:wald-first-order}
\end{equation}
where $\sigma$ is the determinant of the induced horizon metric
and $\mathcal R_{\mathcal H}$ is its intrinsic Ricci scalar.
The cubic interaction contributes only at the next order.

For the equal-spin metric, the horizon area and intrinsic
curvature are
\begin{equation}
    A_{\mathcal H}=2\pi^2 R_h^3\sqrt{W_h},
    \qquad
    \mathcal R_{\mathcal H}
    =\frac{2(4-W_h)}{R_h^2},
\end{equation}
where $R_h=\sqrt{m}\,\rho_h$ and $W_h$ denotes $W$ evaluated
at the horizon. Expanding the area includes both the correction
$s$ to $W$ and the horizon displacement,
\begin{equation}
    \frac{A_{\mathcal H}-A_0}{A_0}
    =
    \frac{\alpha}{m}
    \left[
        \frac{3\delta\rho_+}{\rho_+}
        +\frac{s+\delta\rho_+ W_0'}{2W_0}
    \right]_{\rho=\rho_+}
    +\mathcal O(\epsilon^2),
\end{equation}
with $A_0$ the reference horizon area and primes denoting
derivatives with respect to $\rho$. Substituting the metric
correction and using
$m=\bar r_0^2(1+\bar q)^2/2$ and
$k=2\bar q/(1+\bar q)$ gives
\begin{equation}
    \frac{A_{\mathcal H}-A_0}{4G_N}
    =
    -\frac{\pi^2\alpha\,\bar r_0(1+\bar q)
    (3\bar q^2-2\bar q+3)}
    {2G_N(1-\bar q)}
    +\mathcal O(\epsilon^2).
\end{equation}
The logarithmic terms cancel between the correction to $W$ and the horizon displacement in this area variation.

The explicit Gauss--Bonnet contribution can be evaluated on the
background horizon because it already carries a factor of
$\alpha$. Using $R_+^2=\bar r_0^2(1+\bar q)$ and
$W_0(R_+)=1+\bar q$, we obtain
\begin{equation}
    \frac{\alpha}{2G_N}
    \int_{\mathcal H_0} d^3x\,\sqrt{\sigma_0}\,
    \mathcal R_{\mathcal H_0}
    =
    \frac{2\pi^2\alpha\,\bar r_0(1+\bar q)(3-\bar q)}{G_N}.
\end{equation}
Adding the two contributions yields
\begin{equation}
    \left.\Delta S_{\mathrm{Wald}}\right|_{M,J_i}
    =
    \frac{\pi^2\alpha\,\bar r_0(1+\bar q)
    (\bar q^2-14\bar q+9)}
    {2G_N(1-\bar q)}
    +\mathcal O(\epsilon^2).
    \label{eq:wald-fixed-charge-check}
\end{equation}
This coincides with the term linear in $\alpha$ in the entropy
difference obtained from the Gibbs potential at the same mass
and angular momenta.

\bibliography{SSTwAS}

\end{document}

%% file: coefficient_table.tex
\begin{table}[!htbp]
\centering
\small
\renewcommand{\arraystretch}{1.1}
\setlength{\tabcolsep}{4pt}
\caption{All 48 coefficients of the first-order perturbation normalized
to preserve the ADM mass and both equal angular momenta. The coefficients
are functions of $k$, with $0<k<1$; this dependence is suppressed in the
first column. They refer to the seed before the parameter shifts that
fix $T$ and $\Omega$. The static limit must be taken in the complete
metric functions, where the apparent inverse powers of $k$ cancel.}
\label{tab:coefficients1}
\begin{tabular*}{\textwidth}{@{\extracolsep{\fill}}clcl@{}}
\hline\hline
Coefficient & Value & Coefficient & Value \\
\hline
$\mathcal A_{-1}$ & $0$ & $\mathcal E_{1}$ & $\frac{-16}{\bigl(k-2\bigr)\,k}$ \\[4pt]
$\mathcal A_{0}$ & $\frac{-32}{\bigl(k-2\bigr)^{2}\,k}$ & $\mathcal E_{2}$ & $\frac{4\,\bigl(-30+15\,k-16\,k^{2}+6\,k^{3}\bigr)}{3\,\bigl(k-2\bigr)}$ \\[4pt]
$\mathcal A_{1}$ & $\frac{-16\,\bigl(-14-5\,k+4\,k^{2}\bigr)}{3\,\bigl(k-2\bigr)^{2}\,k}$ & $\mathcal E_{3}$ & $\frac{16\,\bigl(-1-6\,k+3\,k^{2}\bigr)}{3}$ \\[4pt]
$\mathcal A_{2}$ & $\frac{4\,\bigl(16+34\,k+7\,k^{2}-16\,k^{3}+6\,k^{4}\bigr)}{3\,\bigl(k-2\bigr)\,k}$ & $\mathcal E_{4}$ & $\frac{16\,\bigl(k-2\bigr)\,k}{3}$ \\[4pt]
$\mathcal A_{3}$ & $\frac{8\,\bigl(7-4\,k+2\,k^{2}\bigr)}{3}$ & $\mathcal E_{5}$ & $\frac{16\,\bigl(k-2\bigr)^{2}\,k^{2}}{3}$ \\[4pt]
$\mathcal A_{4}$ & $\frac{-16\,\bigl(k-2\bigr)\,k}{3}$ & $\mathcal G_{-1}$ & $\frac{-32}{\bigl(k-2\bigr)^{2}\,k^{2}}$ \\[4pt]
$\mathcal A_{5}$ & $0$ & $\mathcal G_{0}$ & $\frac{32}{\bigl(k-2\bigr)^{2}\,k^{2}}$ \\[4pt]
$\mathcal B_{-1}$ & $\frac{-32}{\bigl(k-2\bigr)^{2}\,k^{2}}$ & $\mathcal G_{1}$ & $\frac{-32}{\bigl(k-2\bigr)\,k}$ \\[4pt]
$\mathcal B_{0}$ & $\frac{-32\,\bigl(-7-4\,k+2\,k^{2}\bigr)}{3\,\bigl(k-2\bigr)^{2}\,k^{2}}$ & $\mathcal G_{2}$ & $\frac{-32\,\bigl(-1-4\,k+2\,k^{2}\bigr)}{3\,\bigl(k-2\bigr)\,k}$ \\[4pt]
$\mathcal B_{1}$ & $\frac{32\,\bigl(-4-10\,k+5\,k^{2}\bigr)}{3\,\bigl(k-2\bigr)^{2}\,k^{2}}$ & $\mathcal G_{3}$ & $0$ \\[4pt]
$\mathcal B_{2}$ & $\frac{-32}{\bigl(k-2\bigr)\,k}$ & $\mathcal I_{-4}$ & $0$ \\[4pt]
$\mathcal C_{-1}$ & $0$ & $\mathcal I_{-3}$ & $\frac{-16}{\bigl(k-2\bigr)^{2}\,k}$ \\[4pt]
$\mathcal C_{0}$ & $\frac{32}{\bigl(k-2\bigr)^{2}\,k}$ & $\mathcal I_{-2}$ & $\frac{-8\,\bigl(-8+k\bigr)}{\bigl(k-2\bigr)^{2}\,k}$ \\[4pt]
$\mathcal C_{1}$ & $\frac{16\,\bigl(-6+k\bigr)}{\bigl(k-2\bigr)^{2}\,k}$ & $\mathcal I_{-1}$ & $\frac{16\,\bigl(4+3\,k\bigr)}{3\,\bigl(k-2\bigr)\,k}$ \\[4pt]
$\mathcal C_{2}$ & $\frac{4\,\bigl(-16-30\,k+15\,k^{2}-16\,k^{3}+6\,k^{4}\bigr)}{3\,\bigl(k-2\bigr)\,k}$ & $\mathcal I_{0}$ & $\frac{-4\,\bigl(-14+13\,k-16\,k^{2}+6\,k^{3}\bigr)}{3\,\bigl(k-2\bigr)}$ \\[4pt]
$\mathcal C_{3}$ & $8\,\bigl(1-4\,k+2\,k^{2}\bigr)$ & $\mathcal I_{1}$ & $\frac{-64\,\bigl(k-2\bigr)\,k}{3}$ \\[4pt]
$\mathcal C_{4}$ & $-16\,\bigl(k-2\bigr)\,k$ & $\mathcal I_{2}$ & $0$ \\[4pt]
$\mathcal C_{5}$ & $\frac{-32\,\bigl(k-2\bigr)^{2}\,k^{2}}{3}$ & $\mathcal I_{3}$ & $0$ \\[4pt]
$\mathcal D_{-1}$ & $\frac{32}{\bigl(k-2\bigr)^{2}\,k^{2}}$ & $\mathcal M_{-4}$ & $\frac{-16}{\bigl(k-2\bigr)^{2}\,k^{2}}$ \\[4pt]
$\mathcal D_{0}$ & $\frac{-96}{\bigl(k-2\bigr)^{2}\,k^{2}}$ & $\mathcal M_{-3}$ & $\frac{64}{\bigl(k-2\bigr)^{2}\,k^{2}}$ \\[4pt]
$\mathcal D_{1}$ & $\frac{-32\,\bigl(-4-10\,k+5\,k^{2}\bigr)}{3\,\bigl(k-2\bigr)^{2}\,k^{2}}$ & $\mathcal M_{-2}$ & $\frac{64\,\bigl(-2-2\,k+k^{2}\bigr)}{3\,\bigl(k-2\bigr)^{2}\,k^{2}}$ \\[4pt]
$\mathcal D_{2}$ & $\frac{-32\,\bigl(-1-4\,k+2\,k^{2}\bigr)}{3\,\bigl(k-2\bigr)\,k}$ & $\mathcal M_{-1}$ & $0$ \\[4pt]
$\mathcal E_{-1}$ & $0$ & $\mathcal M_{0}$ & $16$ \\[4pt]
$\mathcal E_{0}$ & $\frac{-32}{\bigl(k-2\bigr)^{2}\,k}$ & $\mathcal M_{1}$ & $0$ \\[4pt]
\hline\hline
\end{tabular*}
\end{table}